# Effects of solute concentrations on kinetic pathways in Ni-Al-Cr alloys

Chris Booth-Morrison[1], Jessica Weninger[1], Chantal K. Sudbrack[1], Zugang Mao[1], Ronald D. Noebe[2], David N. Seidman[1,3,*]

[1]Department of Materials Science and Engineering, Northwestern University, 2220 Campus Drive, Evanston, Illinois, 60208-3108

[2]NASA Glenn Research Center, 21000 Brookpark Rd., Cleveland, Ohio, 44135

[3]Northwestern University Center for Atom-Probe Tomography (NUCAPT), 2220 Campus Drive, Evanston, Illinois, 60208-3108

* Corresponding author: d-seidman@northwestern.edu, Tel: +1-847-491-4391, Fax: +1-847-467-2269

## Abstract

The kinetic pathways resulting from the formation of coherent $L1_2$-ordered γ'-precipitates in the γ-matrix (f.c.c.) of Ni-7.5 Al-8.5 Cr at.% and Ni-5.2 Al-14.2 Cr at.% alloys, aged at 873 K, are investigated by atom-probe tomography (APT) over a range of aging times from 1/6 to 1024 hours; these alloys have approximately the same volume fraction of the γ'-precipitate phase. Quantification of the phase decomposition within the framework of classical nucleation theory reveals that the γ-matrix solid-solution solute supersaturations of both alloys provide the chemical driving force, which acts as the primary determinant of the nucleation behavior. In the coarsening regime, the temporal evolution of the γ'-precipitate average radii and the γ-matrix supersaturations follow the predictions of classical coarsening models, while the temporal evolution of the γ'-precipitate number densities of both alloys do not. APT results are compared to equilibrium calculations of the pertinent solvus lines determined by employing both *Thermo-Calc* and Grand-Canonical Monte Carlo simulation.

## 1. Introduction

The high temperature strength and creep resistance of modern nickel-based superalloys are due to the presence of coherent, elastically hard, L1$_2$-ordered γ'-precipitates in a γ (f.c.c.) nickel-rich, solid-solution matrix [1]. Efforts to improve these properties by process optimization, and to develop reliable life-prediction techniques, have created a demand for a quantitative understanding of the kinetic pathways that lead to phase decomposition at service temperatures up to 1100 °C [2, 3]. The technological importance of commercial nickel-based superalloys has motivated extensive investigations of the phase separation behavior of model alloys by conventional [4-8] and high–resolution [9] transmission electron microscopy (TEM), x-ray analysis [10-12], small-angle and wide angle neutron scattering [13-17], atom-probe field-ion microscopy (APFIM) [18, 19], atom-probe tomography (APT) [20-26], and phase-field modeling [27-36]. Many of these techniques are, however, limited by either, or both, their spatial and analytical resolutions for composition [37], and thus the early stages of phase decomposition are still not well understood. This is particularly true of concentrated multicomponent alloys.

The research of Schmuck et al. [20, 26] and Pareige et al. [23, 24] combined atom-probe tomography and lattice kinetic Monte Carlo (LKMC) simulations to analyze the decomposition of a Ni-Al-Cr solid-solution. A similar approach was applied by Sudbrack et al. [38-42] and Yoon et al. [43-45] for studying Ni-5.2 Al-14.2 Cr at.% aged at 873 K and Ni–10 Al-8.5 Cr at.%, Ni–10 Al-8.5 Cr-2.0 W at.% and Ni–10 Al-8.5 Cr-2.0 Re at.% aged at 1073 K, which decompose via a first-order phase transformation to form a high number density, $N_v(t)$, of nanometer-sized γ'-precipitates. The addition of Cr to the binary Ni-Al system reduces the lattice parameter misfit between the γ'-Ni$_3$(Al$_x$Cr$_{(1-x)}$)-precipitates and the γ-matrix, often leading to γ'-precipitates that are nearly misfit free [5], thereby allowing the γ'-precipitates to remain spherical or spheroidal to fairly large dimensions as aging progresses [10]. As such, model Ni-Al-Cr alloys are amenable to comparison with predictions from the classical theories of precipitate nucleation, growth and coarsening, which often assume a spheroidal precipitate morphology.

The present investigation focuses on comparing the temporal evolution of the nanostructural and compositional properties of alloy (A), Ni-7.5 Al-8.5 Cr at.% at 873 K, with those previously reported for alloy (B), Ni-5.2 Al-14.2 Cr at. % [38-40] at 873 K; all concentrations herein are in atomic percent (at.%) unless otherwise noted. A ternary Ni-Al-Cr phase diagram determined by the Grand Canonical Monte Carlo (GCMC) technique at 873 K, Figure 1 [46], predicts that the values of $\phi^{eq}$ are 17.5 ± 0.5 and 15.1 ± 0.5 for alloys (A) and (B), respectively. Since the predicted equilibrium volume fractions of the γ'-phase, $\phi^{eq}$, for alloys (A) and (B) are similar, approximately 16%, it follows that any differences observed in decomposition behavior are solely due to differences in solute concentrations.

For comparative purposes, the equilibrium phase boundaries determined by the commercial software package *Thermo-Calc* [47], using two databases for nickel-based superalloys due to Saunders [48] and Dupin et al. [49], are superimposed on the GCMC phase diagram. While the generated γ/γ+γ' solvus lines show good agreement, the curvatures of the γ+γ'/γ' phase lines differs for each technique.

FIGURE 1

*1.1 Classical nucleation theory*

The early stages of phase decomposition by nucleation have been studied theoretically in a set of models known as classical nucleation theory (CNT), which have been reviewed extensively in the literature [37, 50-54]. According to CNT, nucleation is governed by a balance between a bulk free energy term, which has both chemical, $\Delta F_{ch}$, and elastic strain energy, $\Delta F_{el}$, components, and an interfacial free energy term, $\sigma^{\gamma/\gamma'}$, associated with the formation of a γ-matrix/γ'-precipitate heterophase interface; F is the Helmholtz free energy. Thus, the expression for the net reversible work required for the formation of a spherical nucleus, $W_R$, as a function of nucleus radius, $R$, is given by:

$$W_R = (\Delta F_{ch} + \Delta F_{el})\frac{4\pi}{3}R^3 + 4\pi R^2 \sigma^{\gamma/\gamma'} . \qquad (1)$$

According to CNT, the net reversible work acts as a nucleation barrier that nuclei must surmount in order to achieve a critical nucleus radius, $R^*$, having a probability of 0.5 or greater for further growth. The critical net reversible work, $W_R^*$, required for the formation of a critical spherical nucleus is expressed as:

$$W_R^* = \frac{16\pi}{3}\frac{\sigma^{\gamma'/\gamma\,3}}{(\Delta F_{ch} + \Delta F_{el})^2} ; \qquad (2)$$

and $R^*$ is given by:

$$R^* = \frac{2\sigma^{\gamma'/\gamma}}{-(\Delta F_{ch} + \Delta F_{el})}. \qquad (3)$$

For nucleation to occur $(\Delta F_{ch} + \Delta F_{el})$ must be negative. From CNT, the stationary-state nucleation current, $J^{st}$, which is the number of nuclei formed per unit time per unit volume, is given by:

$$J^{st} = Z\beta^* N_0 \exp(\frac{-W_R^*}{k_B T}); \qquad (4)$$

where $Z$, the Zeldovich factor, accounts for the dissolution of supercritical clusters, $\beta$ is a kinetic coefficient describing the rate of condensation of single atoms on the critical nuclei, $N_0$ is the total number of possible nucleation sites per unit volume, taken to be the volume density of lattice points, $k_B$ is Boltzmann's constant and T is the absolute temperature in degrees Kelvin. In their review of nucleation kinetics results for binary alloys, Aaronson and Legoues [55] note that while there exists some experimental evidence to support the correctness of CNT, the nucleation currents predicted by the extant theories are often several orders of magnitude smaller than those measured experimentally, which is most likely due to the presence of precursor clusters that form between the solutionizing and aging treatments. Such precursor clustering has been detected in both alloys (A) [38] and (B), and thus it is anticipated that our calculated values of $J^{st}$ will be smaller than those measured by APT.

*1.2 Coarsening theory*

The first comprehensive mean-field treatment of Ostwald ripening [56], due to Lifshitz and Slyozov [57] and Wagner [58], known as the LSW model, is limited to dilute *binary* alloys with spatially-fixed spherical precipitates whose initial compositions are equal to their equilibrium values. The LSW model for a binary alloy assumes: (i) no elastic interactions among precipitates, thereby limiting the precipitate volume fraction to zero; (ii) precipitates have a spherical morphology; (iii) coarsening occurs in a stress-free matrix; (iv) the precipitate diffusion fields do not overlap; (v) dilute solid-solution theory obtains; (vi) the linearized version of the Gibbs–Thomson equation is valid; (vii) coarsening occurs by the evaporation-condensation mechanism; and (viii) precipitates coarsen with a fixed chemical composition, which is the equilibrium composition. These requirements are

4          6/26/2007

highly restrictive and difficult to meet in practice, and while experimental evidence exists to support the prediction of the time dependency of the mean precipitate radius, <R(t)>, experimentalists have been unable to achieve the exact stationary-state precipitate size distributions (PSDs) predicted by the LSW model [2, 59, 60]. Researchers have worked to remove the mean-field restrictions of the LSW model, and have developed models based on interprecipitate diffusion that are able to describe systems with finite volume fractions [52, 59, 61]. The formation of coherent, spheroidal, γ'-precipitates with relatively stress-free precipitate/matrix heterophase interfaces in the model Ni-Al-Cr alloys studied herein makes them excellent candidates for comparison of experimental data with the predictions of classical treatments of growth and coarsening for *ternary* alloys.

Umantsev and Olson (UO) [62] were the first to demonstrate that the exponents of the temporal power-laws predicted for binary alloys by LSW-type models are identical for multi-component concentrated alloys, but that the explicit expressions for the rate constants depend on the number of components. Kuehmann and Voorhees (KV) [63] considered isothermal quasi-stationary state coarsening in ternary alloys and developed a model that includes the effects of capillarity on the precipitate composition, such that the matrix and precipitate compositions can deviate locally from their equilibrium thermodynamic values. In the quasi-stationary limit at infinite aging time in the KV model, $\delta C_i/\delta t \approx 0$, the exponents of the power-law temporal dependencies for <R(t)>, $N_v(t)$, and the γ-matrix supersaturation, $\Delta C_i^\gamma(t)$, of each solute species *i*, are:

$$<R(t)>^3 - <R(t_0)>^3 = K_{KV}(t-t_0); \qquad (5)$$

$$N_v(t)^{-1} - N_v(t_0)^{-1} = \frac{4.74 K_{KV}}{\phi^{eq}}(t-t_0); \text{ and} \qquad (6)$$

$$\Delta C_i^\gamma(t) = <C_i^{\gamma,ff}(t)> - C_i^{\gamma,eq}(\infty) = \kappa_{i,KV}^\gamma t^{-1/3}; \qquad (7)$$

where $K_{KV}$, and $\kappa_{i,KV}^\gamma$ are the coarsening rate constants for <R(t)> and $\Delta C_i^\gamma(t)$, respectively; <R(t_0)> is the average precipitate radius and $N_v(t_0)$ is the precipitate number density at the *onset* of quasi-stationary coarsening, at time $t_0$. The quantity $\Delta C_i^\gamma(t)$ is denoted a supersaturation and is the difference between the concentration in the far-field γ-matrix, $<C_i^{\gamma,ff}(t)>$, and the equilibrium γ-matrix solute-solubility, $C_i^{\gamma,eq}(\infty)$. The quantity $C_i^{\gamma,eq}(\infty)$ needs to be calculated or determined experimentally as it is not available for Ni-Al-Cr multicomponent alloys at 873 K.

Atom-probe tomography of the nanostructures of alloys (A) and (B) provides an in-depth look at the compositional and nanostructural evolution of the γ'-precipitate phase as it evolves. In this article, the decomposition of the γ-matrix phase, from the earliest stages of solute-rich γ'-nuclei formation, to the subsequent growth and coarsening of γ'-precipitates, is accessed within the framework of classical nucleation, growth, and coarsening theories. The effects of varying the solute concentrations on the temporal evolution of Ni-Al-Cr alloys are determined to provide a more quantitative understanding of the kinetic pathways that lead to phase separation, and the achievement of the equilibrium compositions of both phases. We demonstrate that the kinetic pathways to achieve equilibrium for these alloys are completely different, even though the final volume fractions of the γ'-phase are approximately equal.

**2. Experimental**

High-purity constituent elements (99.97 Ni wt. %, 99.98 Al wt. %, and 99.99 Cr wt. %) were induction-melted and chill cast in a 19 mm diameter copper mold under an Ar atmosphere. The overall compositions of the two alloys were determined by inductively coupled plasma (ICP) atomic-emission spectroscopy, which yielded average atomic compositions of 83.87 Ni-7.56 Al- 8.56 Cr and 80.52 Ni-5.24 Al-14.24 Cr for alloys (A) and (B), respectively. Chemical homogeneity of the cast ingots was achieved by annealing at 1300ºC in the γ-phase field for 20 hours. Next, the ingots were held in the γ–phase field at 850ºC for 3 hours to reduce the concentration of quenched-in vacancies, and then water quenched and sectioned. Ingot sections were then aged at 873 K under flowing argon for times ranging from 1/6 to 1024 h, and microtip specimens were prepared from each of the aged sections for study by atom probe tomography. We performed voltage-pulsed APT with a conventional APT [64, 65] and an Imago Scientific Instrument's local-electrode atom-probe (LEAP®) tomograph [66-68] in the Northwestern University Center for Atom-Probe Tomography (NUCAPT). APT data collection was performed at a specimen temperature of 40.0 ± 0.3 K, a voltage pulse fraction (pulse voltage/steady-state direct current voltage) of 19%, a pulse frequency of 1.5 kHz (conventional APT) or 200 kHz (LEAP® tomograph), and a background gauge pressure of < 6.7 x $10^{-8}$ Pa (5 x $10^{-10}$ torr). The average detection rate in the area of analysis ranged from 0.011 to 0.015 ions per pulse for conventional APT and from 0.04 to 0.08 ions per pulse for the LEAP® tomographic analyses. Conventional APT data were visualized and analyzed with APEX a12 software, the successor of ADAM [69], while LEAP® data was analyzed employing the IVAS® 3.0 software program (Imago Scientific Instruments, Madison, Wisconsin). The γ/γ' interfaces were delineated with 10.5 at.% and 9 at.% Al isoconcentration surfaces generated with efficient sampling procedures [70], for alloys (A) and (B), respectively, and in-depth compositional information was obtained with the proximity histogram method [71]. Further experimental and analytical details for alloy (B) can be found elsewhere [39, 40], the same procedures were employed for alloy (A). The commercial software package *Thermo-Calc* [72] was used to estimate the values of $\phi^{eq}$, $C_i^{\gamma,eq}(\infty)$ and the equilibrium γ'-precipitate composition of each solute species $i$, $C_i^{\gamma',eq}(\infty)$, for Ni-7.5 Al-8.5 Cr and Ni-5.2 Al-14.2 Cr at a pressure of 1 atmosphere, using data bases for nickel-based superalloys due to Saunders [48] and Dupin et al. [49].

### 3. Results

*3.1 Morphological development*

Nanometer-sized spheroidal γ'-precipitates are detected in alloys (A) and (B) over the full range of aging times, from 1/6 to 1024 hours. The temporal evolution of the morphology of alloy (A) is shown in a series of APT micrographs in Figure 2, which can be compared with a similar series for alloy (B) in reference [39]. Figure 3 is a projection of a 25 x 25 x 25 nm$^3$ subset of an APT analysis of alloy (A) aged to 1024 h, showing a spheroidal γ'-precipitate of radius ca. 9 nm, delineated by a dark 10.5% Al isoconcentration surface. Atomic planes are clearly visible within the γ'-precipitate, and the value of the interplanar spacing is 0.26 ± 0.03 nm, suggesting {110}-type planes. This APT image demonstrates that the γ'-precipitates remain spheroidal for aging times as long as 1024 h at an aging temperature of 873 K, as confirmed by the TEM micrograph in Figure 4.

FIGURE 2

FIGURE 3

FIGURE 4

There is evidence of γ'-precipitate coagulation and coalescence in alloys (A) and (B), characterized by the formation of necks that interconnect the γ'-precipitates and exhibit L1$_2$-type ordering. Interconnected γ'-precipitates are first detected by APT in both alloys after 1/4 h of aging, which coincides with the end of the quasi-stationary-state nucleation regime. After 1/4 h, the fraction of coagulating and coalescing γ'-precipitates, $f$, is 15 ± 4% and 9 ± 3%, for alloys (A) and (B), respectively. From Figure 5, the maximum value of $f$ of 18 ± 4% for alloy (A) occurs at an aging time of 1 h, while the maximum for alloy (B) of 30 ± 4% occurs at an aging time of 4 h.

FIGURE 5

*3.2 Temporal evolution of the nanostructural properties of γ'-precipitates*

Figure 6 provides a quantitative description of the temporal evolution of the γ'-precipitate volume fraction, $\phi$, and the quantities $<R(t)>$ and $N_v(t)$, for alloys (A) and (B). The γ'-precipitate nanostructural properties determined by APT analysis for alloy (A), are summarized in Table 1, while reference [39] contains details for alloy (B). The standard errors, $\sigma$, for all quantities are calculated based on counting statistics and reconstruction scaling errors using standard error propagation methods [73], and represent one standard deviation from the mean. The two Ni-Al-Cr alloys were designed to have similar values of $\phi^{eq}$ of approximately 16%, and the values of $\phi$ for alloy (A) and (B) are statistically indistinguishable over the full range of aging times, from 1/6 to 1024 h. It is worth noting that over the range of aging times studied, the values of $<R(t)>$ of alloy (A) are, on average, 32 ± 6% larger than those of alloy (B). From Figure 6, the temporal evolution of alloys (A) and (B) can be divided into three regimes: (i) quasi-stationary-state γ'-precipitate nucleation at early aging times; followed by (ii) concomitant precipitate nucleation and growth, and finally (iii) concurrent growth and coarsening once the maximum value of $N_v(t)$ is achieved.

FIGURE 6, TABLE 1

*3.2.2 Nucleation and growth of γ'-precipitates*

Phase decomposition in alloys (A) and (B) begins with the formation of solute-rich nuclei that grow to form stable γ'-precipitates. Precipitates are first detected by APT in alloys (A) and (B) after 1/6 h of aging. Alloy (A) forms (2.6 ± 1.4) x 10$^{23}$ γ'-precipitates m$^{-3}$ with an $<R(t = 1/6\ h)>$ value of 0.90 ± 0.32 nm, corresponding to a $\phi$ value of 0.31 ± 0.11 %, after 1/6 h of aging. For the same aging time, Alloy (B) forms $(3.6 \pm 1.3) \times 10^{23}$ γ'-precipitates m$^{-3}$ with an $<R(t = 1/6\ h)>$ value of $0.74 \pm 0.24$ nm, accounting for a $\phi$ value of $0.11 \pm 0.04$ %. The nucleation of stable γ'-precipitates in alloy (A) for aging times between 1/6 and 1/4 h results in a sharp linear slope of the $N_v(t)$ profile, $(5.4 \pm 1.5) \times 10^{21}$ m$^{-3}$ s$^{-1}$, while the value of $<R(t)>$ increases only slightly from $0.90 \pm 0.32$ nm to $1.00 \pm 0.11$ nm, which are the same within statistical error. This slope for $N_v(t)$, although based on two experimental data points, is taken to be an estimate of the quasi-stationary-state nucleation current of γ'-precipitates, $J^{st}$. In contrast, alloy (B) undergoes nucleation of stable γ'-precipitates with an essentially constant $<R(t)>$ value of $0.75 \pm 0.24$ nm for aging times less than 1/4 h. The value of $J^{st}$ for alloy (B) is estimated to be $(5.9 \pm 1.7) \times 10^{21}$ m$^{-3}$ s$^{-1}$, which is statistically indistinguishable from the value of $J^{st}$ of $(5.4 \pm 1.5) \times 10^{21}$ m$^{-3}$ s$^{-1}$ measured for alloy (A).

At an aging time of 1/4 h, both alloys (A) and (B) enter a regime of concomitant γ'-precipitate nucleation and growth, which results in steadily increasing $\phi$ and $<R(t)>$ values, and a maximum value for $N_v(t)$. The peak value in $N_v(t)$ of $(2.21 \pm 0.64) \times 10^{24}$ m$^{-3}$ for alloy (A) occurs at 1 h, while for alloy (B), the peak value in $N_v(t)$ of $(3.2 \pm 0.6) \times 10^{24}$ m$^{-3}$ is achieved at 4 hours. We note importantly that the prediction of the temporal dependence of $<R(t)>$ of $t^{1/2}$ for diffusion-limited growth [37, 74, 75], is observed in neither of the Ni-Al-Cr alloys studied, contrary to the results for an earlier APT study of an alloy whose composition is essentially identical to that of alloy (B) [20].

*3.2.3 Growth and coarsening of γ'-precipitates*

Beyond the peak in $N_v(t)$, alloys (A) and (B) enter a quasi-stationary-state of growth and coarsening characterized by a steady diminution of $N_v(t)$ and increasing values of $\phi$ and $<R(t)>$. Beyond 1 h, the quantity $N_v(t)$ for alloy (A) displays a temporal dependence of $t^{-0.42 \pm 0.03}$, which differs significantly from the predicted value of $t^{-1}$ from the UO-KV models. This is not surprising, since the quantity $\phi$ continues to evolve temporally, implying that the system has not achieved a quasi-stationary-state [76]. During coarsening in alloy (B), the quantity $N_v(t)$ displays a temporal dependence of $t^{-0.67 \pm 0.01}$, which also differs from the predicted value of $t^{-1}$, although to a lesser extent. The values of $\phi$ for the two alloys increase steadily in this regime, reaching values of $\phi$ of $16.0 \pm 5.7$ % and $15.6 \pm 6.4$ %, for alloys (A) and (B), respectively, which are statistically indistinguishable. In the growth and coarsening regime, the quantity $<R(t)>$ displays a temporal dependence of $t^{0.29 \pm 0.03}$ for alloy (A) and $t^{0.29 \pm 0.05}$ for alloy (B), which both agree approximately, but not exactly, with the predicted value of $t^{1/3}$, indicating that the phase transformation is primarily diffusion-limited.

From Figure 5, the maximum value for $f$, after 1 h of aging, of 18 ± 4% for alloy (A) coincides with the minimum value of the average edge-to-edge interprecipitate spacing, $<\lambda_{e\text{-}e}>$, of 7.0 ± 2.5 nm and the peak value of $N_v(t)$ of $(2.21 \pm 0.64) \times 10^{24}$ m$^{-3}$. For alloy (B), the peak value of $N_v(t)$, at an aging time of 4 h, is $(3.2 \pm 0.6) \times 10^{24}$ m$^{-3}$, corresponding to a minimum value of $<\lambda_{e\text{-}e}>$ of 5.9 ± 0.8 nm at a maximum value of $f$ of 30 ± 4%. The quantity $<\lambda_{e\text{-}e}>$ is calculated using Equation (8), which assumes a regular simple cubic array of γ'-precipitates [39]:

$$<\lambda_{e-e}> = 2\left[\left(\frac{4}{3}\pi \cdot N_v(t)\right)^{-1/3} - <R(t)>\right]. \quad (8)$$

*3.3 Temporal evolution of the compositions of the γ and γ'-phases*

The compositions of the γ-matrix and the γ'-precipitate phases of alloys (A) and (B) continue to evolve temporally, as the γ-matrix becomes enriched in Ni and Cr and depleted in Al. Figure 7 shows the trajectory of the compositional evolution of the γ-matrix and γ'-precipitate phases on a partial Ni-Al-Cr ternary phase diagram at 873 K. The KV model predicts that the trajectory of the γ-matrix phase during coarsening lies along the tie-line of the alloy, while the trajectory of the γ'-precipitate phase does not, due to the capillary effect on precipitate composition. The trajectories of the γ-matrix phase and the γ'-precipitate phase compositions of alloy (A), determined by APT, have slopes of – 2.56 ± 0.12 and 2.86 ± 0.32, respectively, while the slope of the tie-line, the ratio of the difference in elemental partitioning, is estimated to be -3.50 ± 0.6. Slopes of -1.1 ± 0.29 and 1.21 ± 0.34 are estimated for the γ-matrix and γ'-precipitate phases, respectively, from the APT data for alloy (B), while the slope of the tie-line is -1.53 ± 0.5. From these results, it is absolutely clear that the trajectories of the composition of the γ'-precipitate phases in alloys (A) and (B) do not lie along the equilibrium tie line of the respective alloys. The trajectories of the composition of the γ-matrix phase of alloys (A) and (B) do, however, lie approximately along the tie-lines, although the magnitude of the slope of the experimental compositional trajectory is 27% smaller than that predicted by the KV model for alloy (A) and is 29% smaller than the predicted slope for alloy (B).

FIGURE 7

During the nucleation stage, solute-rich γ'-nuclei form with large values for the Al and Cr supersaturations. Compositional measurements of the first detectable γ'-nuclei measured by APT reflect a solute-supersaturated stoichiometry of $Ni_{2.80\pm0.11}(Al_{0.93\pm0.12}Cr_{0.27\pm0.08})$ at an $<R(t = 1/6\ h)>$ value of 0.90 ± 0.32 nm for alloy (A) and $Ni_{2.85\pm0.12}(Al_{0.76\pm0.11}Cr_{0.39\pm0.08})$ with an $<R(t = 1/6\ h)>$ value of 0.74 ± 0.24 nm for alloy (B). As phase decomposition progresses beyond nucleation, the magnitude of the values of $\Delta C_i^\gamma(t)$ decrease asymptotically toward a value of zero as the equilibrium phase compositions are approached. The equilibrium γ-matrix and γ'-precipitate compositions are extrapolated by fitting the measured concentrations from the quasi-stationary coarsening regime to Equation (7). The γ'-precipitate phase of alloy (A) is estimated to contain 17.82 ± 0.15 Al and 5.85 ± 0.12 Cr, while the γ-matrix has a composition of 5.42 ± 0.09 Al and 9.39 ± 0.09 Cr at infinite time. The γ'-precipitate phase composition of alloy (B) is estimated to be 16.69 ± 0.40 Al and 6.77 ± 0.30 Cr, while the equilibrium composition of the γ-matrix phase is 3.13 ± 0.08 Al and 15.61 ± 0.18 Cr. The equilibrium stoichiometry of the γ'-precipitates is determined to be $Ni_{3.05\pm0.02}(Al_{0.71\pm0.02}Cr_{0.23\pm0.01})$ for alloy (A) and $Ni_{3.06\pm0.02}(Al_{0.67\pm0.02}Cr_{0.27\pm0.01})$ for alloy (B). Summaries of the equilibrium phase compositions of alloy (A) determined by APT are presented in Table 2, and agree with results obtained from *Thermo-Calc* and GCMC simulation. Reference [39] contains a similar table for alloy (B).

TABLE 2

The partitioning behavior of the elements can be determined quantitatively by calculating the partitioning ratio, $K_i^{\gamma/\gamma'}$, defined as ratio of the concentration of an element *i* in the γ'-precipitates to the concentration of the same element in the γ-matrix. Figure 8 demonstrates that alloys (A) and (B) both exhibit partitioning of Al to the γ'-precipitates and of Ni and Cr to the γ-matrix. Partitioning is more pronounced in alloy (B) than (A), as the smaller Al and the larger Cr concentrations of this alloy result in a smaller solubility of Al, and a larger solubility of Cr in the γ-matrix, respectively.

FIGURE 8

The lever rule is applied to the equilibrium phase compositions for both alloys to estimate a $\phi^{eq}$ value of 16.4 ± 0.6 % for alloy (A), and 15.7 ± 0.7 % for alloy (B). The $\phi$ values of 16.0 ± 5.7 % and 15.6 ± 6.4 % determined by APT at an aging time of 1024 h for alloy (A) and (B), are within experimental error of the $\phi^{eq}$ values estimated by the lever rule. The *Thermo-Calc* software package yields $\phi^{eq}$ values of 16.7 % and 14.9 % for alloy (A) according to the Saunders and Dupin et al. databases, respectively, which are approximately close to the experimental value of 16.4 ± 0.6%. Alloy (B) is predicted to achieve $\phi^{eq}$ values of 12.83% and 12.34% for the same databases, which are both smaller than the experimentally determined value of 15.7 ± 0.7%. Summaries of the $\phi^{eq}$ values determined by APT are presented in Table 3, and agree with results obtained from *Thermo-Calc* and GCMC modeling to different degrees.

TABLE 3

The values of the solid-solution supersaturations of alloys (A) and (B) are calculated based on the equilibrium phase compositions from APT data. The initial solid-solution supersaturation values are estimated to be 2.08 ± 0.02 Al and -0.89 ± 0.01 Cr, for alloy (A), and 2.09 ± 0.04 Al and -1.46 ± 0.02 Cr, for alloy (B). Figure 9 exhibits the temporal evolution of the Al and Cr γ-matrix supersaturation values for both alloys. The formation of γ'-nuclei during the early stages of aging results in a decrease in the magnitude of the values of $\Delta C_i^\gamma(t)$, which in turn causes the slowing and eventual termination of γ'-precipitate nucleation. Beyond the aging time corresponding to the peak γ'-precipitate number density, 1 h for alloy (A), and 4 h for alloy (B), the diminution of the $\Delta C_i^\gamma(t)$ values approximately follows the $t^{-1/3}$ prediction of the UO-KV models. From Figure 9, alloy (A) demonstrates a temporal dependence of $t^{-0.32 \pm 0.03}$ for $\Delta C_{Al}^\gamma(t)$ and $t^{-0.29 \pm 0.04}$ for $\Delta C_{Cr}^\gamma(t)$, and alloy (B) exhibits a dependence of $t^{-0.33 \pm 0.04}$ for $\Delta C_{Al}^\gamma(t)$ and $t^{-0.34 \pm 0.07}$ for $\Delta C_{Cr}^\gamma(t)$. The supersaturation values of the γ'-precipitate phases, $\Delta C_i^{\gamma'}(t)$, of alloys (A) and (B) are a reflection of their alloy composition, as the magnitude of $\Delta C_i^{\gamma'}(t)$ is greater in alloy (A), Ni-7.5 Al-8.5 Cr, which contains more Al, than in alloy (B), Ni-5.2 Al-14.2 Cr, while the inverse is true for Cr. From Figure 10, the quantities $\Delta C_{Al}^{\gamma'}(t)$ and $\Delta C_{Cr}^{\gamma'}(t)$ for alloy (A) exhibit temporal dependencies of $t^{-0.34 \pm 0.04}$ and $t^{-0.32 \pm 0.05}$ respectively, which are equivalent to the predicted value of $t^{1/3}$. A dependence of $t^{-0.30 \pm 0.05}$ for $\Delta C_{Al}^{\gamma'}(t)$, and $t^{-0.29 \pm 0.07}$ for $\Delta C_{Cr}^{\gamma'}(t)$, are demonstrated for alloy (B).

FIGURES 9 & 10

## 4. Discussion

While the temporal evolution of the morphology and the volume fraction of the γ'-precipitates in alloy (A) Ni-7.5 Al-8.5 Cr and alloy (B) Ni-5.2 Al-14.2 Cr are similar, the nanostructural and compositional results for the two alloys exhibit significant differences. These differences can be attributed to the effects of solute concentrations on the kinetic pathways involved in phase decomposition.

*4.1 Effects of solute concentration on nucleation behavior*

The nucleation behavior observed by APT for alloys (A) and (B) can be compared with the predictions of CNT using Equations (2-4), given the values of $\sigma^{\gamma/\gamma'}$, $\Delta F_{ch}$, and $\Delta F_{el}$ for both alloys. The values of $\sigma^{\gamma/\gamma'}$ are estimated as first shown by Ardell for a binary alloy [77, 78], and later for a ternary alloy by Marquis and Seidman [79], and as applied to alloy (B) by Sudbrack et al. [40]. The relationship for $\sigma^{\gamma/\gamma'}$ in a nonideal, nondilute ternary alloy consisting of a γ-matrix and a γ'-precipitate phase with a finite volume fraction of the γ'-phase is given by:

$$\sigma^{\gamma/\gamma'} = \frac{(K_{KV})^{1/3} \kappa_{i,KV}^{\gamma}}{2 V_m^{\gamma'} p_i} (p_{Al}^2 G_{Al,Al}^{\gamma} + p_{Al} p_{Cr} G_{Al,Cr}^{\gamma} + p_{Cr}^2 G_{Cr,Cr}^{\gamma}) ; \qquad (9)$$

where $K_{KV}$ and $\kappa_{i,KV}^{\gamma}$ are the rate constants for the quantities $<R(t)>$ and $\Delta C_i^{\gamma}(t)$, respectively, from the KV coarsening model, $V_m^{\gamma'}$ is the molar volume of the γ'-precipitate phase, calculated to be 6.7584 x $10^{-6}$ m³mol⁻¹, $p_i$ is the magnitude of the partitioning as defined by $p_i = C_i^{\gamma',eq}(\infty) - C_i^{\gamma,eq}(\infty)$, and $G_{i,j}^{\gamma}$ is shorthand notation for the partial derivatives of the molar Gibbs free-energy of γ-matrix phase with respect to the solute species $i$ and $j$. The quantities $K_{KV}$ and $\kappa_{i,KV}^{\gamma}$ are determined by fitting the experimental APT data to Equations (5) and (7) from the KV coarsening model. For alloy (A), $K_{KV}$ is $(1.84 \pm 0.43) \times 10^{-31}$ m³s⁻¹ and $\kappa_{Al,KV}^{\gamma}$ and $\kappa_{Cri,KV}^{\gamma}$ are $0.18 \pm 0.05$ at.fr. s$^{1/3}$ and $-0.06 \pm 0.01$ at.fr. s$^{1/3}$, while a value of $K_{KV}$ of $(8.8 \pm 3.3) \times 10^{-32}$ m³s⁻¹, and values of $\kappa_{Al,KV}^{\gamma}$ and $\kappa_{Cri,KV}^{\gamma}$ of $0.19 \pm 0.02$ at.fr. s$^{1/3}$ and $-0.14 \pm 0.05$ at.fr. s$^{1/3}$, are found for alloy (B). For the general case described by nonideal and nodilute solution theory, the values of $G^{\gamma}_{i,j}$ may be calculated using *Thermo-Calc* employing the extant databases for nickel-based superalloys [48, 49]. As noted by Sudbrack et al. [40], the calculations of $G^{\gamma}_{i,j}$ from *Thermo-Calc* predict a more highly-curved free energy surface than ideal solution theory with respect to all solute species combinations. The *Thermo-Calc* assessments take into account the excess free-energies of mixing and the magnitudes of $G^{\gamma}_{i,j}$ are 1.5 to 13 times larger than those for ideal solution theory for alloys (A) and (B) as shown in Table 5 for alloy (A). A similar table is given by Sudbrack et al. [40] for the values of $G^{\gamma}_{i,j}$ for alloy (B).

Fortuitously [40], the *Thermo-Calc* and ideal solution theory assessments of $G^{\gamma}_{i,j}$ yield approximately the same value for $\sigma^{\gamma/\gamma'}$ for alloy (B) of 22-23 ± 7 mJ m$^{-2}$ [40]. For alloy (A), the values for $\sigma^{\gamma/\gamma'}$ calculated from the ideal solution theory of 14-16 ± 3 mJ m$^{-2}$ are smaller than those from the *Thermo-Calc* assessment of $G^{\gamma}_{i,j}$ of 23-25 ± 6 mJ m$^{-2}$, Table 5. For the purposes of CNT, values of $\sigma^{\gamma/\gamma'}$ of 24 ± 6 mJ m$^{-2}$, for alloy (A), and 22.5 ± 7 mJ m$^{-2}$, for alloy (B) are used, the averages of the values generated from the *Thermo-Calc* assessments of $G^{\gamma}_{i,j}$. Additionally, estimates of 36.3 ± 3.8 mJ m$^{-2}$ and 35.7 ± 1.7 mJ m$^{-2}$ are obtained for alloys (A) and (B) from first principles calculations in the framework of density functional theory (DFT) and the local density approximation (LDA), employing ultrasoft Vanderbilt potentials (US-PP), using the VASP (Vienna *Ab-initio* simulation package) code [80]. It is noted that the values of $\sigma^{\gamma/\gamma'}$ from the first principles calculations are larger than those determined experimentally because the first principles calculations assume a sharp γ/γ' interface and are performed at 0 K, and therefore do not include entropic effects, whereas the experimental estimates of $\sigma^{\gamma/\gamma'}$ are for experimentally diffuse γ/γ' interfaces at 873 K and are free energies because they include entropic effects.

TABLE 4

A value of $\sigma^{\gamma/\gamma'}$ of 12.5 mJ m$^{-2}$ was previously determined for a ternary Ni-5.2 Al-14.8 Cr alloy aged at 873 K by Schmuck et al. [20]. The value of $\sigma^{\gamma/\gamma'}$ for this alloy, whose overall composition is very close to that of alloy (B) Ni-5.2 Al-14.2 Cr, was determined assuming the LSW model for *binary* alloys. We reanalyzed their data according to the method developed by Marquis and Seidman for *ternary* alloys, and values of 20-21 ± 5 mJ m$^{-2}$ are estimated for $\sigma^{\gamma/\gamma'}$, in good agreement with the values of $\sigma^{\gamma/\gamma'}$ estimated for alloys (A) and (B); additional details are presented in the Appendix. Gleiter and Hornbogen [81] estimated a value of 13.5 mJ m$^{-2}$ for a Ni-5.4 Al-18.7 Cr alloy aged at 750 ºC, though they also applied binary LSW theory to a ternary alloy, and their results did not include concentration data, making recalculation of their value of $\sigma^{\gamma/\gamma'}$ impossible by the method developed by Marquis and Seidman for ternary alloys. Baldan [2] provides a review of literature values of $\sigma^{\gamma/\gamma'}$ for several different Ni-Al and Ni-Al-Cr systems over a range of aging temperatures, which need to be evaluated in detail in light of our work.

TABLE 5

The values of the chemical driving force of alloys (A) and (B) are estimated as proposed by Lupis [82] and applied by Schmuck et al. [20], from thermodynamic data taken from the commercial software package *Thermo-Calc* [72] and the extant nickel-based superalloy databases [48, 49]. The values of $\Delta F_{ch}$ for alloys (A) and (B) are estimated to be -443.4 J mol$^{-1}$ and -556.8 J mol$^{-1}$, respectively. The values of $\Delta F_{el}$ for the two alloys are estimated using [83]:

$$\Delta F_{el} = \frac{2S^{\gamma} B^{\gamma'} (V_a^{\gamma'} - V_a^{\gamma})^2}{(3B^{\gamma'} + 4S^{\gamma})V_a^{\gamma'}} ;\qquad [9]$$

where $S^{\gamma}$ is the shear modulus of the γ-matrix phase, $B^{\gamma'}$ is the bulk modulus of the γ'-precipitate phase, and $V_a^{\gamma}$ and $V_a^{\gamma'}$ are the atomic volumes of the γ-matrix and γ'-precipitate phases, respectively. No elastic constants are available for this alloy, therefore the value of $S^{\gamma}$ of 100.9 GPa, of a similar alloy, Ni-12.69 Al at 873 K [84] is employed, while the value of $B^{\gamma'}$ is taken to be 175 GPa [85]. The lattice parameters for the equilibrium phases in alloy (A) at 873 K, are estimated to be 0.3554 ± 0.0001 nm and 0.3544 ± 0.0001 nm for the γ' and γ-phases, respectively, based on room-temperature x-ray diffraction measurements on similar Ni-Al-Cr alloys [2,7]. These lattice parameter values result in a near-zero estimate of the lattice misfit of 0.0027 ± 0.0004 for alloy (A). The lattice parameter misfit for alloy (B) is estimated to be 0.0006 ± 0.0004 [39]. Substituting these values into Equation [8] for $\Delta F_{el}$ yields values of 17.1 J mol$^{-1}$ for alloy (A) and 0.732 J mol$^{-1}$ for alloy (B). The larger Cr concentration in alloy (B) is responsible for a smaller value of the lattice parameter misfit and therefore a smaller elastic strain energy. The quantity $\Delta F_{el}$ is also estimated by a simpler technique due to Eshelby [86], which yields values of 18.9 J mol$^{-1}$ and 0.93 J mol$^{-1}$ for alloys (A) and (B), respectively. The high degree of coherency of the γ'-precipitates in these alloys is such that the bulk component of the driving force for nucleation from Equation (1) is dominated by the $\Delta F_{ch}$ term, as $\Delta F_{el}$ is only 3.9% of the value of $\Delta F_{ch}$ for alloy (A) and 0.1% for alloy (B). As such, experimentally determined differences in the nucleation behavior may be described as due primarily, but not exclusively, to differing values of $\Delta F_{ch}$. It is noted that the intrinsic diffusivity of Al in these Ni-Al-Cr alloys is significantly larger than the diffusivity of Cr, as shown by Sudbrack et al. for alloy (B), where $D_{AlAl}^{fcc}$ is 22 x $10^{-21}$ $m^2s^{-1}$, while $D_{CrCr}^{fcc}$ is 7.0 x $10^{-21}$ $m^2s^{-1}$ [40]. As such, the kinetics of the early stages of the phase transformation of alloy (A) Ni-7.5 Al-8.5 Cr are expected to be faster than those involved in the decomposition of alloy (B) Ni-5.2 Al-14.2 Cr, as a result of the larger Al concentration of alloy (A) and the larger Cr concentration of alloy (B).

The value of $W_R^*$ for alloy (A), Equation (2), is estimated to be 35.2 kJ $mol^{-1}$ or 0.365 eV $atom^{-1}$, while $R^*$ is estimated, Equation (3), to be 0.76 nm. The value of $W_R^*$ for alloy (B) is found to be 17.0 kJ $mol^{-1}$ or 0.177 eV $atom^{-1}$, and $R^*$ is calculated to be 0.55 nm. The nucleation currents for the two alloys are estimated, Equation (7), to be 5.4 x $10^{23}$ $m^{-3}$ $s^{-1}$ for alloy (A) and 6.1 x $10^{24}$ $m^{-3}$ $s^{-1}$ for alloy (B). The values of $\sigma^{\gamma/\gamma'}$, $\Delta F_{ch}$, $\Delta F_{el}$, $W_R^*$, $R^*$, and $J^{st}$ for both alloys are summarized in Table 6.

TABLE 6

The predictions of the value of $R^*$ from CNT are verified experimentally for both alloys, as the first nucleating precipitates detected consistently by APT at an aging time of 1/6 h have $<R(t)>$ values of 0.9 ± 0.32 nm for alloy (A) and 0.74 ± 0.24 nm for alloy (B), which are both slightly greater than the calculated $R^*$ estimates of 0.76 and 0.55 nm, respectively. The predicted value of $J^{st}$ of 5.4 x $10^{23}$ $m^{-3}$ $s^{-1}$ for alloy (A) is two orders of magnitude greater than the experimental value of (5.4 ± 1.5) x $10^{21}$ $m^{-3}$ $s^{-1}$. The calculated value of $J^{st}$ for alloy (B) of 6.1 x $10^{24}$ $m^{-3}s^{-1}$ is three orders of magnitude greater than the experimentally measured value of $J^{st}$ of (5.9 ± 1.7) x $10^{21}$ $m^{-3}$ $s^{-1}$. Given the evidence of precursor clustering in these Ni-Al-Cr alloys [38], it is surprising that the experimentally determined values of $J^{st}$ are orders of magnitude less than the predicted values. Xiao and Haasen [9] performed a similar comparison of experimentally determined nucleation currents with those predicted by CNT for a binary Ni-12.0 Al at.% alloy aged at 773 K and found that the predicted value of $J^{st}$ was a factor of 500 larger than the measured value of $J^{st}$. They attributed this discrepancy to the sensitivity of the predicted value of $J^{st}$ to the value of $R^*$. Xiao and Haasen also pointed out that the predicted nucleation currents are likely an overestimate due to the assumption that the value of the total number of nucleation sites per volume is equal to the volume density of lattice points, which is a commonly made assumption that is not necessarily correct. Given that the experimentally determined values of $J^{st}$ are measured from two experimental data points, and that the detailed kinetics involved in the formation of γ'-nuclei in these ternary systems are not completely understood, further analysis is not instructive. The nucleation behavior of Ni-Al-Cr alloys certainly warrants future research, given the technological importance of these systems.

The formation of stable, growing nuclei at early aging times causes a decrease in the values of $\Delta C_i^\gamma(t)$, and results in a decrease in, and the eventual termination of, nucleation. By an aging time of 1 h, the values of $\Delta C_i^\gamma(t)$ for alloy (A) decrease from the solid-solution values of 2.08 ± 0.02 Al and -0.89 ± 0.01 Cr to 1.21 ± 0.02 Al and -0.52 ± 0.04 Cr. This decrease in the magnitude of the $\Delta C_i^\gamma(t)$ values is reflected in a 50.1% decrease in the quantity $\Delta F_{ch}$ to a value of -220.9 J mol$^{-1}$, and therefore a decrease in the predicted value of $J^{st}$ of seven orders of magnitude, after only 1 h of aging. After 4 h of aging in alloy (B), the values of $\Delta C_i^\gamma(t)$ decrease from 2.09 ± 0.04 Al and -1.46 Cr, to 0.84 ± 0.04 Al and -0.59 ± 0.09 Cr, while the value of $\Delta F_{ch}$ decreases 51.2% to a value of -271.4 J mol$^{-1}$. This decrease in the quantity $\Delta F_{ch}$ results in a decrease in the predicted value of $J^{st}$ of only three orders of magnitude for alloy (B). From these estimates, and our APT nanostructural results, it appears that the larger initial value of the chemical driving force in alloy (B) sustains a significant nucleation current for aging times as long as 4 h, whereas the nucleation current decreases significantly in alloy (A) after only 1 h.

*4.1 Effects of Solute Concentration on the Coarsening Behavior*

As alloys (A) and (B) coarsen, the quantities $\Delta C_i^\gamma(t)$ and $\Delta C_i^{\gamma'}(t)$ evolve temporally as approximately $t^{-1/3}$, and $<R(t)>$ grows as approximately $t^{1/3}$, as predicted by the UO-KV models. The experimentally measured temporal dependencies of $N_v(t)$ of $t^{-0.42 \pm 0.03}$ and $t^{-0.67 \pm 0.01}$, for alloys (A) and (B), differ significantly, however, from the $t^{-1}$ prediction of both models. The magnitudes of the quantity $\Delta C_i^\gamma(t)$ for both alloys remain non-zero at an aging time of 1024 h, thus neither alloy has achieved a true stationary-state, which is a basic assumption of classical LSW Ostwald ripening behavior. The UO-KV models, however, assume a quasi-stationary state, which *may* have been achieved for alloys (A) and (B).

When Ostwald ripening is considered to be a diffusion-limited process, it is limited by the characteristic length over which diffusion can occur, taken as the average edge-to-edge interprecipitate spacing in these systems. It is important to note that as $N_v(t)$ decreases with increasing time, $<R(t)>$ increases and the value of $<\lambda_{e-e}>$ is concomitantly increasing. The time required to reach stationary-state coarsening, $t_c$, may be estimated employing [87]:

$$t_c = \frac{K_{KV}^6 K_{\lambda_{e-e}}^6}{216 D^3};$$ [10]

where $K_{KV}$ is the coarsening rate constant for $<R(t)>$ as defined in section 1.2, $D$ is the diffusivity of the least mobile atomic species in the alloy, taken to be the intrinsic diffusivity of Cr, which has a value of 7.0 x $10^{-21}$ m$^2$ s$^{-1}$ [40], and $K_{\lambda_{e-e}}$ is the constant in the radial dependence of the interprecipitate spacing as defined by Nembach [88], and is a function of the value of $\phi^{eq}$. Estimates of $t_c$ for alloy (A) from Equation (10) range from 1200 to 1800 days, while the predicted values of $t_c$ for alloy (B) range from 330 to 475 days. The estimates of $t_c$ for both alloys are well beyond the longest aging time studied (1024 h, or 42.7 days). Our inability to achieve stationary-state coarsening explains the continuously increasing values of $\phi$ and the non-zero values of $\Delta C_i^\gamma(t)$ and $\Delta C_i^{\gamma'}(t)$ at an aging time of 1024 h, which may explain the deviation from the $t^{-1}$ prediction for the temporal evolution of $N_v(t)$. The values of $t_c$ for alloy (A) are significantly longer than those estimated for alloy (B), an indication that the coarsening kinetics are slower in alloy (A). Thus the rate of the diminution of $N_v(t)$ is significantly slower in alloy (A) ($N_v(t) \sim t^{-0.42 \pm 0.03}$) than in alloy (B) ($N_v(t) \sim t^{-0.67 \pm 0.01}$).

A recent study by Mao et. al [89] combined APT and LKMC to study the role of the precipitation diffusion mechanism on the early-stage precipitate morphology of alloy (B). It is shown that the long-range solute-vacancy binding energies (out to fourth nearest-neighbor distance) strongly affect the γ'-precipitate coagulation and coalescence process, which occurs abundantly at early stages. Coagulation and coalescence are shown to result from the overlap of nonequilibrium concentration profiles surrounding γ'-precipitates that give rise to nonequilibrium diffuse interfaces. The concentration profiles associated with the interfacial regions between γ'-precipitates are spread over distances significantly larger than that of the equilibrium interfacial thickness. This is due to specific couplings between the diffusion fluxes of the constituents elements toward and away from γ'-precipitates, which are a result of the finite vacancy-solute binding energies. From this analysis, coagulation and coalescence are therefore more likely when there is a strong attractive solute-vacancy binding energy and when $<\lambda_{e-e}>$ has a minimum value, which occurs when $N_v(t)$ has its maximum value, as evidenced by the larger $f$ values observed in alloy (B). Additionally, the larger Al concentration in alloy (A) leads to the formation of more highly mobile Al clusters than in alloy (B), which explains why the quantities $N_v(t)$ and $f$ achieve their maximum values and the quantity $<\lambda_{e-e}>$ reaches a minimum value after aging for only 1h in alloy (A), while this condition is reached at 4 h in alloy (B).

## 5. Summary and Conclusions

We present a detailed comparison of the nanostructural and compositional evolution of alloy (A) Ni-7.5 Al-8.5 Cr and alloy (B) Ni-5.2 Al-14.2 Cr, during phase separation at 873 K for aging times ranging from 1/6 to 1024 h, employing atom-probe topography (APT). These alloys are designed to have similar equilibrium γ'-precipitate volume fractions ($\phi$), estimated from APT results to be 16.4 ± 0.6% for alloy (A) and 15.7 ± 0.7% for alloy (B), to study the effects of solute concentration on the kinetic pathways in model Ni-Al-Cr alloys, leading to the following results:

- The morphology of the γ'-precipitate phase in both alloys is found by both APT and TEM to be spheroidal for aging times as long as 1024 h, as a result of a near-zero lattice parameter misfit between the γ-matrix and γ'-precipitate phases. This high degree of coherency makes these alloys amenable to a comparison of APT results to predictions of classical theories of nucleation, growth and coarsening, where the chemical free energy term is dominant. γ'-precipitate coagulation and coalescence is observed and is argued to be a result of the overlap of the nonequilibrium concentration profiles associated with adjacent γ'-precipitates [85].

- After aging to 1/6 h, precipitation of the γ'-precipitate phase is evident for both alloys, as alloy (A) forms nuclei with a stoichiometry of Ni$_{2.80\pm0.11}$(Al$_{0.93\pm0.12}$Cr$_{0.27\pm0.08}$) at a γ'-precipitate number density, $N_v(t = 1/6\ h)$, of $(2.6 \pm 1.4) \times 10^{23}$ m$^{-3}$, a mean radius, $<R(t = 1/6\ h)>$, of 0.90 ± 0.32 nm and a volume fraction, $\phi$, of 0.31 ± 0.11 %. Alloy (B) forms $Ni_{2.85\pm0.12}(Al_{0.76\pm0.11}Cr_{0.39\pm0.08})$ nuclei with a $N_v(t = 1/6\ h)$ value of $(3.6 \pm 1.3) \times 10^{23}$ m$^{-3}$, an $<R(t = 1/6\ h)>$ value of $0.74 \pm 0.24$ nm, and a $\phi$ value of $0.11 \pm 0.04\%$.

- Classical nucleation theory (CNT) is applied to demonstrate that the chemical free energy change for forming a nucleus, $\Delta F_{ch}$, is -443.4 J mol$^{-1}$ for alloy (A) and -556.8 J mol$^{-1}$ for alloy (B), providing the primary driving force for nucleation. As such, the differing solute concentrations of the two alloys are responsible for differences in the nucleation behavior and in the nanostructural and compositional evolution of the alloys as they decompose. The high degree of γ'-precipitate coherency with the γ-matrix in these alloys is such that the elastic strain energy values, $\Delta F_{el}$, estimated to be 17.1 J mol$^{-1}$ for alloy (A), and 0.732 J mol$^{-1}$ for alloy (B), are only a small fraction of the values of $\Delta F_{ch}$, for both alloys.

- Estimates of the γ/γ' interfacial free energy values, $\sigma^{\gamma/\gamma'}$, from the coarsening data obtained by APT yield values of $23\text{-}25 \pm 6$ mJ m$^{-2}$ for alloy (A) and $22\text{-}23 \pm 6$ mJ m$^{-2}$ for alloy (B). These values are in good agreement with our recalculated value of $\sigma^{\gamma/\gamma'}$ for the coarsening data of Schmuck et al. [20] for a ternary Ni-5.2 Al-14.8 Cr alloy aged at 873 K, that gives a value of $20\text{-}21 \pm 5$ mJ m$^{-2}$, which is greater than the value they calculated by applying the LSW coarsening model for a binary alloy to their ternary alloy.

- The predictions of the calculated critical nucleus radius required for nucleation, $R^*$, from CNT, of 0.76 nm and 0.55 nm for alloys (A) and (B), respectively, show reasonable agreement with the average radii of the first precipitates detected by APT at 1/6 h, of $0.90 \pm 0.32$ nm and $0.74 \pm 0.24$ nm. The predicted value of $J^{st}$ of $5.4 \times 10^{23}$ m$^{-3}$ s$^{-1}$ for alloy (A) is two orders of magnitude greater than the experimental value of $(5.4 \pm 1.5) \times 10^{21}$ m$^{-3}$ s$^{-1}$, and the calculated value of $J^{st}$ for alloy (B) of $6.1 \times 10^{24}$ m$^{-3}$s$^{-1}$ is three orders of magnitude greater than the experimentally measured value of $J^{st}$ of $(5.9 \pm 1.7) \times 10^{21}$ m$^{-3}$ s$^{-1}$. This discrepancy may be due to the sensitivity of the predicted value of $J^{st}$ to the value of $R^*$, and due to the assumption that the value of the total number of nucleation sites per volume is equal to the volume density of lattice points which leads to an overestimate of the nucleation current, as suggested, for example, by Xiao and Haasen [9]. Further research is required to measure the value of $J^{st}$ more accurately for ternary Ni-Al-Cr alloys in order to better understand the nucleation kinetics in these concentrated multicomponent alloys. To date there is not a generally accepted theory of nucleation in concentrated multicomponent alloys.

- After 1 h of aging, alloy (A) achieves a peak value of $N_v(t)$ of $(2.21 \pm 0.64) \times 10^{24}$ m$^{-3}$, which coincides with a minimum value of the average interprecipitate edge-to-edge spacing, $<\lambda_{e\text{-}e}>$, of $7.0 \pm 2.5$ nm and a maximum in the fraction of interconnected precipitates, $f$, of $18 \pm 4$ %. At an aging time of 4 h, Alloy (B) achieves a peak value of $N_v(t)$ of $(3.2 \pm 0.6) \times 10^{24}$ m$^{-3}$ at a minimum value of $<\lambda_{e\text{-}e}>$ of $5.9 \pm 0.8$ nm and a maximum value of $f$ of $30 \pm 4$ %. The larger peak value of $N_v(t)$ for alloy (B) is a result of the larger initial solid-solution value of $\Delta F_{ch}$ in alloy (B), which sustains nucleation for longer aging times.

- In the coarsening regime that follows the peak in the value of $N_v(t)$, the quantity $<R(t)>$ displays a temporal dependence of $t^{0.29 \pm 0.03}$ for alloy (A) and $t^{0.29 \pm 0.05}$ for alloy (B), in approximate agreement with the $t^{1/3}$ prediction of the Umantsev and Olson (UO) and Kuehmann and Voorhees (KV) models for isothermal coarsening in ternary alloys. The supersaturation of the γ-matrix phase in Alloy (A) exhibits a temporal dependence of $t^{-0.32 \pm 0.03}$ for Al and $t^{-0.29 \pm 0.04}$ for Cr, while alloy (B) demonstrates a temporal dependence of $t^{-0.33 \pm 0.04}$ for Al and $t^{-0.34 \pm 0.07}$ for Cr, in good agreement with the $t^{1/3}$ prediction of the UO-KV models. The supersaturation values of Al and Cr in the γ'-precipitate phase of Alloy (A) display a temporal dependence of $t^{-0.34 \pm 0.04}$ for Al and $t^{-0.32 \pm 0.05}$ for Cr, while the supersaturation values in alloy (B) demonstrate a temporal dependence of $t^{-0.30 \pm 0.05}$ for Al and $t^{-0.29 \pm 0.07}$ for Cr.

- During coarsening, the diminution of the quantity $N_v(t)$ in alloy (A) shows a temporal dependence of $t^{-0.42 \pm 0.03}$, while alloy (B) exhibits a temporal dependence of $t^{-0.67 \pm 0.01}$. These temporal exponents deviate from the $t^{-1}$ prediction of the UO-KV coarsening models, because neither alloy has achieved stationary-state coarsening, characterized by a constant value of $\phi$, and a zero value of the γ-matrix supersaturation. Estimates of the critical time required for stationary-state coarsening, $t_c$, for alloy (A) range from 1200 to 1800 days, while for alloy (B) they range from 330 to 475 days. The estimates of $t_c$ for both alloys are well beyond the longest aging time we studied, 1024 h, or 42.7 days, and this is most likely the reason that stationary-state coarsening is not achieved.

- The compositional trajectories of the γ-matrix during phase decomposition lie approximately along the tie-lines, while the trajectories of the γ'-precipitate phase do not, as predicted by the KV model for quasi-stationary state isothermal coarsening in ternary alloys. The compositional trajectories of the γ'-precipitate phase do not lie along the tie lines due to the capillary effect on γ'-precipitate composition. The solute solubility in the γ-matrix phase of alloy (A) is determined by APT to be 5.42 ± 0.09 Al and 9.39 ± 0.09 Cr, and 3.13 ± 0.08 Al and 15.61 ± 0.18 Cr for alloy (B). The equilibrium stoichiometry of the γ'-precipitates is determined to be to be $Ni_{3.05±0.02}(Al_{0.71±0.02}Cr_{0.23±0.01})$ for alloy (A) and $Ni_{3.06±0.02}(Al_{0.67±0.02}Cr_{0.27±0.01})$ for alloy (B).

- The results of Grand Canonical Monte Carlo simulations and *Thermo-Calc* calculations are in reasonable agreement with the experimental values of the phase compositions and equilibrium volume fractions of γ'-precipitates determined by APT.

**Acknowledgements**

This research was sponsored by the National Science Foundation (NSF) under grant DMR-0241928. C. Booth-Morrison and C.K. Sudbrack received partial support from Le Fonds Quebecois de la Recherche sur la Nature et les Technologies (FQRNT) and NSF graduate research fellowships, respectively. Jessica Weninger was supported by an NSF REU award when she was an undergraduate at Northwestern University. Atom-probe tomographic measurements were performed at the Northwestern University Center for Atom Probe Tomography (NUCAPT). The LEAP® tomograph was purchased with funding from the NSF-MRI (DMR 0420532, Dr. Charles Boudin, grant officer) and ONR-DURIP (N00014-0400798, Dr. Julie Christodoulou, grant officer) programs. Additionally, the LEAP® tomograph was enhanced in April 2006 with a picosecond laser with funding from the ONR-DURIP (N00014-0610539, J. Christodoulou, grant officer). We extend our gratitude to Dr. Kevin Yoon for his assistance with *Thermo-Calc* and Dr. Georges Martin for discussions.

**Appendix**

Schmuck et al. [20] determined a value for the interfacial free energy, $\sigma^{\gamma/\gamma'}$, of 12.5 mJ m$^{-2}$ for a ternary Ni-5.2 Al-14.8 Cr alloy aged at 873 K. This value is approximately one half the value determined by Sudbrack et al. [39, 40] for a similar alloy Ni-5.2 Al-14.2 Cr aged at 873 K. The Schmuck et al. approach assumes: (i) the LSW model for *binary* alloys applies to their *ternary* system; (ii) the rate constant for the average γ'-precipitate radius from LSW is correct, which it is not [90]; and (iii) that the γ-matrix and γ'-precipitate phase compositions achieved their equilibrium compositions after aging for 64 h, which it does not. We improve on the Schmuck et al. approach by applying a relationship for $\sigma^{\gamma/\gamma'}$ in a nonideal, nondilute ternary alloy as proposed by Marquis and Seidman [79], based on the results of Calderon et al. [90] and the Kuehmann-Voorhees coarsening model [63], and as applied to alloy (B) by Sudbrack et al. [40]. Furthermore, Schmuck et al. found that the γ'-precipitates had compositions very close to their equilibrium values after only 1 h, and had ostensibly achieved their equilibrium compositions after 64 h. Sudbrack et al. [39, 40] find, however, that the γ'-precipitate phase compositions of a similar alloy, alloy (B) Ni-5.2 Al-14.2 Cr, continue to evolve at an aging time of 1024 h at 873 K. Therefore, estimates of the γ'-precipitate equilibrium composition from the limited APT data recorded by Schmuck et al. are inaccurate, and, from Table 7, do not agree with estimates of equilibrium compositions from *Thermo-Calc* for Ni-5.2 Al-14.8 Cr. As a result, we use the equilibrium compositions calculated by *Thermo-Calc*, employing the Saunders database [48], in our estimates of the value of $\sigma^{\gamma/\gamma'}$ from the data of Schmuck et al. A coarsening rate constant, $K_{KV}$, of (2.30 ± 0.13) x $10^{-31}$ m$^3$ s$^{-1}$ and values of $\kappa^{\gamma}_{Al,KV}$ and $\kappa^{\gamma}_{Cr,KV}$ of 0.13 ± 0.01 at.fr. s$^{1/3}$ and -0.08 ± 0.02 at.fr. s$^{1/3}$, respectively, are found from the data of Schmuck et al. These values should be compared to values of $K_{KV}$ of (8.8 ± 3.3) x $10^{-31}$ m$^3$ s$^{-1}$ and of 0.19 ± 0.02 at.fr.s$^{1/3}$ and -0.14 ± 0.05 at.fr.s$^{1/3}$, for $\kappa^{\gamma}_{Al,KV}$ and $\kappa^{\gamma}_{Cr,KV}$ respectively, for alloy (B). The corrected value of $\sigma^{\gamma/\gamma'}$ calculated from the data of Schmuck et al. using Equation (8) is 20-21 ± 5 mJ m$^{-2}$, which is in good agreement with values of 22-23 ± 7 mJ m$^{-2}$ for alloy (B) and 23-25 ± 6 mJ m$^{-2}$ for alloy (A).

TABLE 7

**List of Symbols and Acronyms**

Alloy (A) Ni-7.5 Al-8.5 Cr at.%
Alloy (B) Ni-5.2 Al-14.2 Cr at.%

γ matrix phase
γ' Ni$_3$(Al$_{1-x}$,Cr$_x$)-type precipitated phase

$< C_i^{\gamma,f\!f}(t) >$ far-field γ-matrix concentration of atomic species *i*
$C_i^{\gamma,eq}(\infty)$ equilibrium γ-matrix composition of each atomic species *i*

$C_i^{\gamma',eq}(\infty)$ equilibrium γ'-precipitate composition of each atomic species $i$

$\Delta C_i^{\gamma}(t)$ supersaturation of element i in the γ-matrix phase

$\Delta C_i^{\gamma'}(t)$ supersaturation of element i in the γ'-precipitate phase

$D$ diffusivity

$f$ fraction of interconnected γ'-precipitates

$\Delta F_{ch}$ chemical free energy change on forming a γ'-nucleus

$\Delta F_{el}$ elastic free energy change on forming a γ'-nucleus

$G_{i,j}^{\gamma}$ partial derivatives of the molar Gibbs free energy of the γ-matrix phase

$J^{st}$ quasi-stationary-state nucleation current of γ'-precipitates

$k_B$ Boltzmann's constant

$K_{KV}$ coarsening rate constant for $<R(t)>$ from the KV model

$K_i^{\gamma/\gamma'}$ partitioning ratio of each atomic species $i$

$K_{\lambda_{e-e}}$ constant in the radial dependence of the interprecipitate spacing

$B^{\gamma'}$ bulk modulus of the γ'-precipitate phase

$B^{\gamma}$ bulk modulus of the γ'-precipitate phase

$N_0$ total number of possible nucleation sites per unit volume

$N_v(t)$ number density of γ'-precipitates

$N_v(t_0)$ precipitate number density at the onset of quasi-stationary coarsening

$p_i$ magnitude of the partitioning between two phases

$R$ radius of the γ'-precipitate/γ'-nucleus

$R^*$ radius of the critical γ'-nucleus

$<R(t)>$ mean γ'-precipitate radius

$<R(t_0)>$ average precipitate radius at the onset of quasi-stationary coarsening

$t_0$ time at the onset of quasi-stationary coarsening

$t_c$ time required to reach stationary-state coarsening

$T$ absolute temperature in degrees Kelvin

$V_a^{\gamma}$ atomic volume of the γ-matrix

$V_a^{\gamma'}$ atomic volume of the γ'-precipitate phase

$V_m^{\gamma'}$ molar volume of the γ'-precipitate phase

$W_R$ net reversible work required to form a γ'-nucleus

$W_R^*$ net reversible work required to form a critical nucleus

$Z$ Zeldovich factor

$\beta$ kinetic coefficient describing the rate of condensation of a single atom on the critical nucleus

$\phi$ volume fraction of the γ'-precipitate phase

$\phi^{eq}$ equilibrium volume fraction of the γ'-precipitate phase

$\kappa_{i,KV}^{\gamma}$ coarsening rate constant of the atomic species $i$ for $\Delta C_i^{\gamma}(t)$ from the KV model

$<\lambda_{e-e}>$ mean edge-to-edge interprecipitate spacing

$S^{\gamma}$ shear modulus of the γ-matrix phase

σ one standard deviation from the mean

$\sigma^{\gamma/\gamma'}$ γ-matrix/γ'-precipitate interfacial energy

**References**
[1] Durand-Charre M. The Microstructure of Superalloys. Amsterdam: Gordon and Breach Science, 1997.
[2] Baldan A. J Mater Sci 2002;37:2379-405.
[3] Reed RC. The Superalloys: Fundamentals and Applications New York: Cambridge University Press, 2006.
[4] Hirata T, Kirkwood DH. Acta Metall 1977;25:1425-34.
[5] Taylor A, Floyd W. J Inst Met 1952-1952;81:451-64.
[6] Ardell AJ, Nicholson RB. J Phys Chem Solids 1966;27:1793-804.
[7] Kirkwood DH. Acta Metall 1970;18:563-70.
[8] Chellman DJ, Ardell AJ. Acta Metall 1974;22:577-88.
[9] Xiao SQ, Haasen P. Acta Metall Mater 1991;39:651-59.
[10] Hornbogen E, Roth M. Z Metall 1967;58:842-55.
[11] Royer A, Bastie P, Veron M. Acta Mater 1998;46:5357-68.
[12] Broz P, Svoboda M, Bursik J, Kroupa A, Havrankova J. Mat Sci Eng A-Struct 2002;A325:59-65.
[13] Messoloras S, Stewart RJ. J Appl Crystallogr 1988;21:870-2.
[14] Beddoe R, Haasen P, Kostorz G. Early Stages of Decomposition in Ni-Al Single Crystals Studied by Small-Angle Neutron Scattering. In: Haasen P GV, Wagner R, and Ashby MF, editor. Decomposition of Alloys: The Early Stages. Oxford: Pergamon Press, 1984.
[15] Staron P, Kampmann R. Acta Mater 2000;48:701-12.
[16] Bruno G, Pinto HC. Mater Sci Tech Ser 2003;19:567-72.
[17] Gilles R, Mukherji D, Hoelzel M, Strunz P, Toebbens DM, Barbier B. Acta Mater 2006;54:1307-16.
[18] Wendt H, Haasen P. Acta Metall 1983;31:1649-59.
[19] Vanbakel GPEM, Hariharan K, Seidman DN. Appl Surf Sci 1995;90:95-105.
[20] Schmuck C, Caron P, Hauet A, Blavette D. Philos Mag A 1997;76:527-42.
[21] Chambreland S, Walder A, Blavette D. Acta Metall 1988;36:3205-15.
[22] Miller MK. Micron 2001;32:757-64.
[23] Pareige-Schmuck C, Soisson F, Blavette D. Mat Sci Eng A-Struct 1998;250:99-103.
[24] Pareige C, Soisson F, Martin G, Blavette D. Acta Mater 1999;47:1889-99.
[25] Kitashima T, Ping DH, Harada H, Kobayashi T. J Jpn Inst Met 2006;70:184-87.
[26] Schmuck C, Danoix F, Caron P, Hauet A, Blavette D. Appl Surf Sci 1996;94-5:273-79.
[27] Shen C, Mills MJ, Wang Y. Proc Mater Res Soc Sym 2003;753:309-14.
[28] Zhou SH, Wang Y, Zhu JZ, Wang T, Chen LQ, MacKay RA, Liu Z-K. Superalloys 2004, Proc Int Sym Superalloys 2004:969-75.
[29] Simmons JP, Wen Y, Shen C, Wang YZ. Mat Sci Eng A-Struct 2004;A365:136-43.
[30] Wang J, Osawa M, Yokokawa T, Harada H, Enomoto M. Superalloys 2004, Proc Int Sym Superalloys 2004:933-40.
[31] Chu Z, Chen Z, Wang YX, Lu YL, Li YS. Prog Nat Sci 2005;15:656-60.
[32] Chu Z, Chen Z, Wang YX, Lu YL, Li YS. J Mater Sci Technol 2006;22:315-20.
[33] Wang Y, Banerjee D, Su CC, Khachaturyan AG. Acta Mater 1998;46:2983-3001.
[34] Wen YH, Wang B, Simmons JP, Wang Y. Acta Mater 2006;54:2087-99.
[35] Wen YH, Simmons JP, Shen C, Woodward C, Wang Y. Acta Mater. 2003;51:1123-32.
[36] Grafe U, Bottger B, Tiaden J, Fries SG. Scripta Mater 2000;42:1179-86.
[37] Wagner R, Kampmann R, Voorhees PW. Homogeneous Second-Phase Precipitation. Weinheim: Wiley-VCH, 2001.
[38] Sudbrack CK, Noebe RD, Seidman DN. Phys Rev B 2006;73:212101/1-01/4.

[39]     Sudbrack CK, Yoon KE, Noebe RD, Seidman DN. Acta Mater. 2006;54:3199-210.
[40]     Sudbrack CK, Noebe RD, Seidman DN. Accepted by Acta. Mater. 2006.
[41]     Sudbrack CK, Isheim D, Noebe RD, Jacobson NS, Seidman DN. Microsc Microanal 2004;10:355-65.
[42]     Sudbrack CK, Ziebell TD, Noebe RD, Seidman DN. to be submitted to Acta Mater. 2007.
[43]     Yoon KE, Sudbrack CK, Noebe RD, Seidman DN. Z Metall 2005;96:481-85.
[44]     Yoon KE, Noebe RD, Seidman DN. Acta Mater 2007;55:1145-57.
[45]     Yoon KE, Noebe RD, D.N. S. Acta Mater 2007;55:1159-69.
[46]     Mao Z, Sinnott SB, Martin G, Seidman DN. to be submitted to Acta Materialia 2006.
[47]     Sundman B, Jansson B, Andersson JO. Calphad 1985;9:153-90.
[48]     Saunders N. Superalloys 1996, Proc Int Sym Superalloys 1996:101-10.
[49]     Dupin N, Ansara I, Sundman B. Calphad 2001;25:279-98.
[50]     Martin G. The Theories of Unmixing Kinetics of Solid Solutions. Solid State Phase Transformation in Metals and Alloys. Orsay, France: Les Éditions de Physique, 1978. p.337-406.
[51]     Russell KC. Adv Colloid Interfac 1980;13:205-318.
[52]     Ratke L, Voorhees PW. Growth and Coarsening, Ripening in Material Processing. Berlin: Springer-Verlag, 2002.
[53]     Kashchiev D. Nucleation: Basic Theory and Applications. Oxford: Elsevier Science, 2000.
[54]     Brechet Y, Martin G. C R Phys 2006;7:959-76.
[55]     Aaronson HI, Legoues FK. Metall Trans A 1992;23:1915-45.
[56]     Greenwood GW. Acta Metall. 1956;4:243–48.
[57]     Lifshitz IM, Slyozov VV. J Phys Chem Solids 1961;19:35-50.
[58]     Wagner C. Z Electrochem 1961;65:581-91.
[59]     Baldan A. J Mater Sci 2002;37:2171-202.
[60]     Rowenhorst DJ, Kuang JP, Thornton K, Voorhees PW. Acta Mater 2006;54:2027-39.
[61]     Voorhees PW. Anu Rev Mater Sci 1992;22:197-215.
[62]     Umantsev A, Olson GB. Scripta Metall Mater 1993;29:1135-40.
[63]     Kuehmann CJ, Voorhees PW. Metall Mater Trans A 1996;27A:937-43.
[64]     Blavette D, Deconihout B, Bostel A, Sarrau JM, Bouet M, Menand A. Rev Sci Instrum 1993;64:2911-19.
[65]     Cerezo A, Godfrey TJ, Sijbrandij SJ, Smith GDW, Warren PJ. Rev Sci Instrum 1998;69:49-58.
[66]     Bajikar SS, Larson DJ, Kelly TF, Camus PP. Ultramicroscopy 1996;65:119-29.
[67]     Kelly TF, Camus PP, Larson DJ, Holzman LM, Bajikar SS. Ultramicroscopy 1996;62:29-42.
[68]     Kelly TF, Larson DJ. Mater Charact 2000;44:59-85.
[69]     Hellman O, Vandenbroucke J, du Rivage JB, Seidman DN. Mat Sci Eng A-Struct 2002;327:29-33.
[70]     Hellman OC, Blatz du Rivage J, Seidman DN. Ultramicroscopy 2003;95:199-205.
[71]     Hellman OC, Vandenbroucke JA, Rusing J, Isheim D, Seidman DN. Microsc Microanal 2000;6:437-44.
[72]     Sundman B, Jansson B, Andersson JO. CALPHAD: Computer Coupling of Phase Diagrams and Thermochemistry 1985;9:153-90.
[73]     Parratt LG. Probability and Experimental Errors in Science. New York: John Wiley, 1966.
[74]     Zener C. J. Appl. Phys. 1949;20:950-3.
[75]     Ham FS. J. Appl. Phys. 1959;30:1518-25.
[76]     Ardell AJ. Mat Sci Eng A-Struct 1997;A238:108-20.
[77]     Ardell AJ. Acta Metall 1968;16:511-16.
[78]     Ardell AJ. Acta Metall 1967;15:1772-75.
[79]     Marquis EA, Seidman DN. Acta Mater 2005;53:4259-68.

[80]     Mao Z. Work in Progress.
[81]     Gleiter H, Hornbogen E. Z Metall 1967;58:157-63.
[82]     Lupis CHP. Chemical Thermodynamics of Materials. New York: North-Holland, 1983.
[83]     Christian JW. Theory of Transformations in Metals and Alloys, Part 1. Oxford, U. K.: Pergamon Press, 1975.
[84]     Prikhodko SV, Carnes JD, Isaak DG, Ardell AJ. Scripta Mater. 1997;38:67-72.
[85]     Stassis C, Kayser FX, Loong CK, Arch D. Phys Rev B 1981;24:3048-53.
[86]     Eshelby JD. Proc R Soc London, Ser A 1957;241:376–96.
[87]     Karnesky RA, Martin G, Seidman DN. to be submitted to Scripta. Mat. 2007.
[88]     Nembach E. Particle Strengthening of Metals and Alloys. New York: John Wiley and Sons, 1996.
[89]     Mao Z, Sudbrack CK, Yoon KE, Martin G, Seidman DN. Nature Mater. 2006;6:210-16.
[90]     Calderon HA, Voorhees PW, Murray JL, Kostorz G. Acta Metall Mater 1994;42:991-1000.

Table 1. Temporal evolution of the nanostructural properties of γ'-precipitates [a] determined by APT for Ni-7.5 Al-8.5 Cr aged at 873 K.

| Aging time (h) | $N_{ppt}$ [b] | $<R(t)> \pm \sigma$ (nm) | $N_v(t) \pm \sigma$ (x $10^{24}$ m$^{-3}$) | $\phi \pm \sigma$ (%) | $f \pm \sigma$ (%) | $<\lambda_{e\text{-}e}> \pm \sigma$ (nm) |
|---|---|---|---|---|---|---|
| 1/6 | 8 | 0.90 ± 0.32 | 0.26 ± 0.14 | 0.31 ± 0.11 | ND [c] | 17.7 ± 6.3 |
| 1/4 | 101 | 1.00 ± 0.11 | 1.89 ± 0.59 | 1.36 ± 0.14 | 15 ± 4 | 7.7 ± 2.7 |
| 1 | 70.5 | 1.24 ± 0.12 | 2.21 ± 0.64 | 2.48 ± 0.25 | 18 ± 4 | 7.0 ± 2.5 |
| 4 | 46 | 1.70 ± 0.25 | 1.02 ± 0.31 | 5.98 ± 0.88 | 16 ± 3 | 8.9 ± 3.2 |
| 16 | 42 | 2.80 ± 0.43 | 0.60 ± 0.18 | 9.12 ± 1.4 | 13 ± 3 | 9.1 ± 3.2 |
| 64 | 76.5 | 3.59 ± 0.41 | 0.34 ± 0.10 | 11.8 ± 1.4 | 7.4 ± 1.9 | 10.6 ± 3.7 |
| 256 | 15 | 5.54 ± 1.43 | 0.20 ± 0.05 | 14.6 ± 3.8 | 2.5 ± 0.9 | 10.1 ± 3.6 |
| 1024 | 8 | 8.30 ± 2.93 | 0.13 ± 0.05 | 16.0 ± 5.7 | ND [c] | 7.9 ± 2.8 |

[a] Mean radius of γ'-precipitates, $<R(t)>$, the number density, $N_v(t)$, precipitated volume fraction, $\phi$, fraction of γ'-precipitates interconnected by necks, $f$, average edge-to-edge interprecipitate spacing, $<\lambda_{e\text{-}e}>$, and their standard errors, σ, one standard deviation is reported.

[b] The number of precipitates analyzed, $N_{ppt}$, is smaller than the total number of precipitates intersected during 3DAP microscopy analyses as precipitates intersected by the sample volume contribute 0.5 to this quantity.

[c] ND = not detected.

Table 2. Equilibrium γ'-precipitate and γ-matrix equilibrium concentrations, as determined by atom-probe tomography (APT), Grand Canonical Monte Carlo (GCMC) simulation, and thermodynamic modeling employing *Thermo-Calc* for alloy (A) Ni-7.5 Al-8.5 Cr aged at 873 K.

| Equilibrium composition of γ'-precipitates | Ni (at.%) | Al (at.%) | Cr (at.%) |
| --- | --- | --- | --- |
| Measured by APT at 1024 h: | 76.11 ± 0.09 | 18.02 ± 0.09 | 5.87 ± 0.05 |
| Extrapolated from APT data: | 76.33 ± 0.03 | 17.82 ± 0.04 | 5.85 ± 0.03 |
| Modeled by GCMC simulation [46]: | 76.3 ± 0.5 | 17.8 ± 0.5 | 5.9 ± 0.5 |
| Calculated with *Thermo-Calc* and Saunders database [48]: | 76.40 | 17.79 | 5.81 |
| Calculated with *Thermo-Calc* and Dupin database [49]: | 75.44 | 17.87 | 6.99 |
| Equilibrium composition of γ-matrix | Ni (at.%) | Al (at.%) | Cr (at.%) |
| Measured by APT at 1024 h: | 85.13 ± 0.06 | 5.53 ± 0.07 | 9.34 ± 0.04 |
| Extrapolated from APT data: | 85.19 ± 0.01 | 5.42 ± 0.02 | 9.39 ± 0.01 |
| Modeled by GCMC simulation [46]: | 85.8 ± 0.5 | 5.2 ± 0.5 | 9.0 ± 0.5 |
| Calculated with *Thermo-Calc* and Saunders database [48]: | 85.71 | 5.42 | 8.86 |
| Calculated with *Thermo-Calc* and Dupin database [49]: | 85.3 | 5.99 | 8.70 |

Table 3. Equilibrium γ'-precipitate volume fraction, $\phi^{eq}$, as determined by atom-probe tomography (APT), Grand Canonical Monte Carlo (GCMC) simulation, and thermodynamic modeling in *Thermo-Calc* for alloy (A) Ni-7.5 Al-8.5 Cr, and alloy (B) Ni-5.2 Al-14.2 Cr, aged at 873 K.

| Technique used to estimate $\phi^{eq}$ | (A) Ni-7.5 Al-8.5 Cr | (B) Ni-5.2 Al-14.2 Cr |
|---|---|---|
| Determined by lever rule calculation with phase compositions measured by APT: | 16.4 ± 0.6 | 15.7 ± 0.7 |
| Measured by APT at 1024 h: | 16.0 ± 5.7 | 15.6 ± 6.4 |
| Modeled by GCMC simulation [46]: | 17.5 ± 0.5 | 15.1 ± 0.5 |
| Calculated with *Thermo-Calc* and Saunders database [48]: | 16.69 | 12.83 |
| Calculated with *Thermo-Calc* and Dupin database [49]: | 14.90 | 12.34 |

Table 4. Curvatures in the molar Gibbs free-energy surface of the γ-matrix phase evaluated at the equilibrium composition with respect to components $i$ and $j$, $G^{\gamma}_{i,j}$, obtained from ideal solution theory and *Thermo-Calc* thermodynamic assessments for alloy (A) Ni-7.5 Al-8.5 Cr aged at 873 K.

| $G^{\gamma}_{i,j}$ | Ideal Solution Theory (J mol$^{-1}$) | Saunders database [48] (J mol$^{-1}$) | Dupin et al. database [49] (J mol$^{-1}$) |
| --- | --- | --- | --- |
| $G^{\gamma}_{Al,Al}$ | 142,441.7 | 271,392.5 | 306,697.8 |
| $G^{\gamma}_{Cr,Cr}$ | 85,821.1 | 166,601.4 | 166,708.5 |
| $G^{\gamma}_{Al,Cr}$ | 8,521.4 | 112,567.3 | 140,244.9 |

Table 5. Free-energy of the γ/γ' interfaces, $\sigma^{\gamma/\gamma'}$, at 873 K in Ni–7.5 Al–8.5 Cr calculated from the experimental values of the coarsening rate constants for the average precipitate radius and the supersaturation of solute species *i* employing Equation (9) with solution thermodynamics described by the ideal solution and *Thermo-Calc* databases in Table 4.

| Thermodynamic models | $\sigma_{Al}^{\gamma/\gamma'}$ (mJ m$^{-2}$) | $\sigma_{Cr}^{\gamma/\gamma'}$ (mJ m$^{-2}$) | $\sigma^{\gamma/\gamma'}$ (mJ m$^{-2}$) |
|---|---|---|---|
| Ideal Solution Theory | 14.2 ± 3.1 | 16.0 ± 3.6 | 15.1 ± 3.4 |
| Saunders database [48] | 23.6 ± 4.9 | 26.6 ± 6.2 | 25.1 ± 5.5 |
| Dupin et al. database [49] | 21.6 ± 4.0 | 24.4 ± 6.1 | 23.0 ± 5.1 |

Table 6. The interfacial free energy, $\sigma^{\gamma/\gamma'}$, the chemical free energy, $\Delta F_{ch}$, and the elastic strain energy, $\Delta F_{el}$, components of the driving force for nucleation used to estimate the net reversible work, $W_R^*$, required for the formation of critical nuclei of size, $R^*$, and the nucleation current, $J^{st}$, according to classical nucleation theory.

| Alloy | $\sigma^{\gamma/\gamma'}$ (mJ m$^{-2}$) | $\Delta F_{ch}$ (J mol$^{-1}$) | $\Delta F_{el}$ (J mol$^{-1}$) | $W_R^*$ (kJ mol$^{-1}$) | $R^*$ (nm) | $J^{st}$ (m$^{-3}$ s$^{-1}$) |
|---|---|---|---|---|---|---|
| (A) Ni-7.5 Al-8.5 Cr | 24.0 ± 6 | -443.4 | 17.1 | 35.2 | 0.76 | 5.4 x 10$^{23}$ |
| (B) Ni-5.2 Al-14.2 Cr | 22.5 ± 7 | -556.8 | 0.732 | 17.0 | 0.55 | 6.1 x 10$^{24}$ |

Table 7. Equilibrium γ'-precipitate and γ-matrix equilibrium concentrations, as determined by atom-probe tomography (APT) and *Thermo-Calc* for Ni-5.2 Al-14.8 Cr aged at 873 K from Schmuck et al. [20].

| Equilibrium composition of γ'-precipitates | Ni (at.%) | Al (at.%) | Cr (at.%) |
| --- | --- | --- | --- |
| Measured by APT at 64 h [20]: | 74 ± 2 | 18 ± 1 | 7.6 ± 0.8 |
| Extrapolated from APT data [20]: | 74.2 ± 0.9 | 18.4 ± 0.8 | 7.4 ± 0.5 |
| Calculated with *Thermo-Calc* and Saunders database [48]: | 74.91 | 16.08 | 9.01 |
| Calculated with *Thermo-Calc* and Dupin database [49]: | 75.79 | 14.44 | 9.76 |
| Equilibrium composition of γ-matrix | Ni (at.%) | Al (at.%) | Cr (at.%) |
| Measured by APT at 64 h [20]: | 80.2 ± 0.3 | 4 ± 0.2 | 15.8 ± 0.2 |
| Extrapolated from APT data [20]: | 80.3 ± 0.2 | 3.94 ± 0.08 | 15.8 ± 0.2 |
| Calculated with *Thermo-Calc* and Saunders database [48] : | 80.80 | 3.50 | 15.70 |
| Calculated with *Thermo-Calc* and Dupin database [49]: | 80.65 | 3.77 | 15.58 |

**Figure Captions**

Figure 1. A partial ternary phase diagram of the Ni-Al-Cr system at 873 K calculated using the Grand Canonical Monte Carlo simulation technique [46], showing the proximity of both alloys (A) Ni-7.5 Al- 8.5 Cr and (B) Ni-5.2 Al-14.2 Cr to the ($\gamma + \gamma'$) / $\gamma$ solvus line. Equilibrium solvus curves determined by *Thermo-Calc* [72], using databases for nickel-based superalloys due to Dupin et al. [49] and Saunders [48], are superimposed on the GCMC phase diagram for comparative purposes.

Figure 2. The temporal evolution of the $\gamma'$-precipitate nanostructure in Ni-7.5 Al-8.5 Cr aged at 873 K is shown in a series of APT parallelepipeds. The parallelepipeds are 10 x 10 x 25 nm$^3$ subsets of the analyzed volume and contain ca. 125,000 atoms. The $\gamma/\gamma'$ interfaces are delineated in gray with 10.5% Al isoconcentration surfaces.

Figure 3. A subset of an APT micrograph of Ni-7.5 Al- 8.5 Cr aged at 873 K for 1024 hours, containing 350,000 atoms, with the Ni and Cr atoms omitted for clarity. A $\gamma'$-precipitate with a radius ca. 9 nm is delineated by the dark 10.5% Al isoconcentration surface, and shows {110}-type planes with an interplanar spacing of 0.26 ± 0.03 nm.

Figure 4. A centered superlattice reflection dark–field image of spheroidal Ni$_3$(Al$_x$Cr$_{(1-x)}$) $\gamma'$-precipitates, **g** = [111] is the operating reflection, for a Ni-7.5 Al-8.5 Cr sample aged for 1024 h at 873 K. Image recorded near the [011] zone axis.

Figure 5. The temporal evolution of the value of the fraction of $\gamma'$-precipitates interconnected by necks, $f$, and the minimal average interprecipitate edge-to-edge spacing, $<\lambda_{e-e}>$. The maximum value of $f$ 18 ± 4% corresponds to the minimum value of $<\lambda_{e-e}>$ of 7 ± 2 nm at an aging time of 1 h for alloy (A) Ni-7.5 Al-8.5 Cr. For alloy (B) Ni-5.2 Al-14.2 Cr, the minimum value of $<\lambda_{e-e}>$ of 5.9 ± 0.8 nm and the maximum value of $f$ of 30 ± 4% coincide at an aging time of 4 h

Figure 6. The temporal evolution of the γ'-precipitate volume fraction, $\phi$, number density, $N_v(t)$, and mean radius, $<R(t)>$, for (A) Ni-7.5 Al-8.5 Cr and (B) Ni-5.2 Al-14.2 Cr alloys aged at 873 K as determined by APT microscopy. The quantity $<R(t)>$ is proportional to $t^{1/3}$ as predicted by the Umantsev and Olson (UO) and Kuehmann and Voorhees (KV) models for isothermal coarsening in ternary alloys. The temporal dependence of the diminution of the quantity $N_v(t)$ deviates from the $t^{-1}$ prediction of the UO-KV models for both alloys.

Figure 7. The compositional trajectories of the temporal evolution of the γ-matrix and γ'-precipitate phases of alloys (A) Ni-7.5 Al-8.5 Cr and (B) Ni-5.2 Al-14.2 Cr displayed on a partial Ni-Al-Cr ternary phase diagram at 873 K. The tie-lines are drawn between the equilibrium phase compositions determined by extrapolation of APT concentration data to infinite time. The trajectories of the γ-matrix phases in alloys (A) and (B) lie approximately on the experimental tie-lines. The trajectories of the γ'-precipitate phases do not lie along the tie-line, as a result of the capillary effect [63].

Figure 8. The partitioning ratio, $K_i^{\gamma/\gamma'}$, of Al and Cr demonstrates that both alloys exhibit partitioning of Al to the γ'-precipitates, and Cr to the γ-matrix. Partitioning is more pronounced in alloy (B), where the smaller Al and larger Cr concentration results in a smaller value of the γ-matrix Al solubility, and a larger γ-matrix Cr solubility.

Figure 9. The magnitude of the values of the supersaturations, $\Delta C_i^{\gamma}(t)$, of Al and Cr in the γ-matrix are smaller for alloy (A) Ni-7.5 Al-8.5 Cr than for alloy (B) Ni-5.2 Al-14.2 Cr. The magnitude of the $\Delta C_i^{\gamma}(t)$ values decrease as $t^{-1/3}$ for both alloys, as predicted by the Umantsev and Olson (UO) and Kuehmann and Voorhees (KV) models for isothermal coarsening in ternary alloys.

Figure 10. The supersaturation values of the γ'-precipitates, $\Delta C_i^{\gamma'}(t)$, reflect the chemical compositions of the two alloys, as the value of $\Delta C_{Al}^{\gamma'}(t)$ is larger in alloy (A) Ni-7.5 Al- 8.5 Cr, which contains more Al, than in alloy (B) Ni-5.2 Al-14.2 Cr, while the inverse is true for Cr. The values of $\Delta C_i^{\gamma'}(t)$ decrease as approximately $t^{-1/3}$ for both alloys.

FIGURE 1

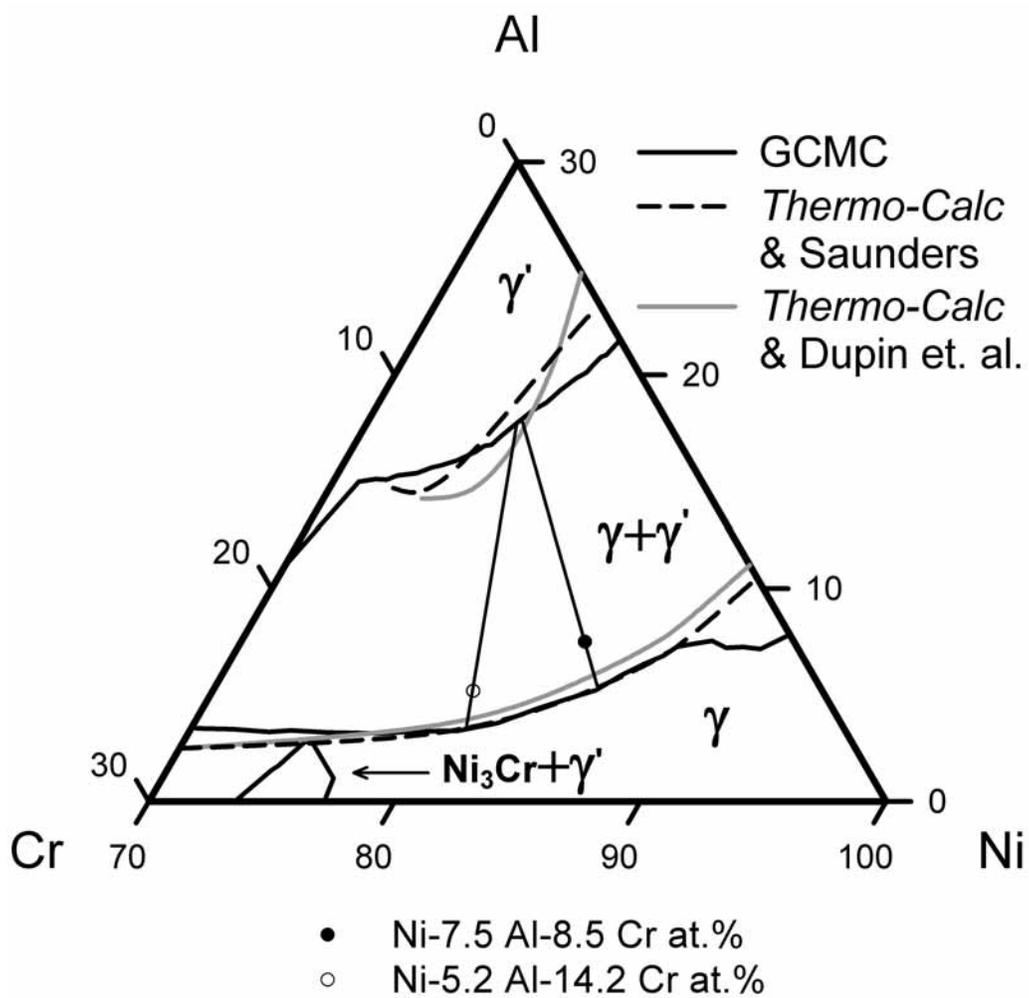

- ● Ni-7.5 Al-8.5 Cr at.%
- ○ Ni-5.2 Al-14.2 Cr at.%

FIGURE 2

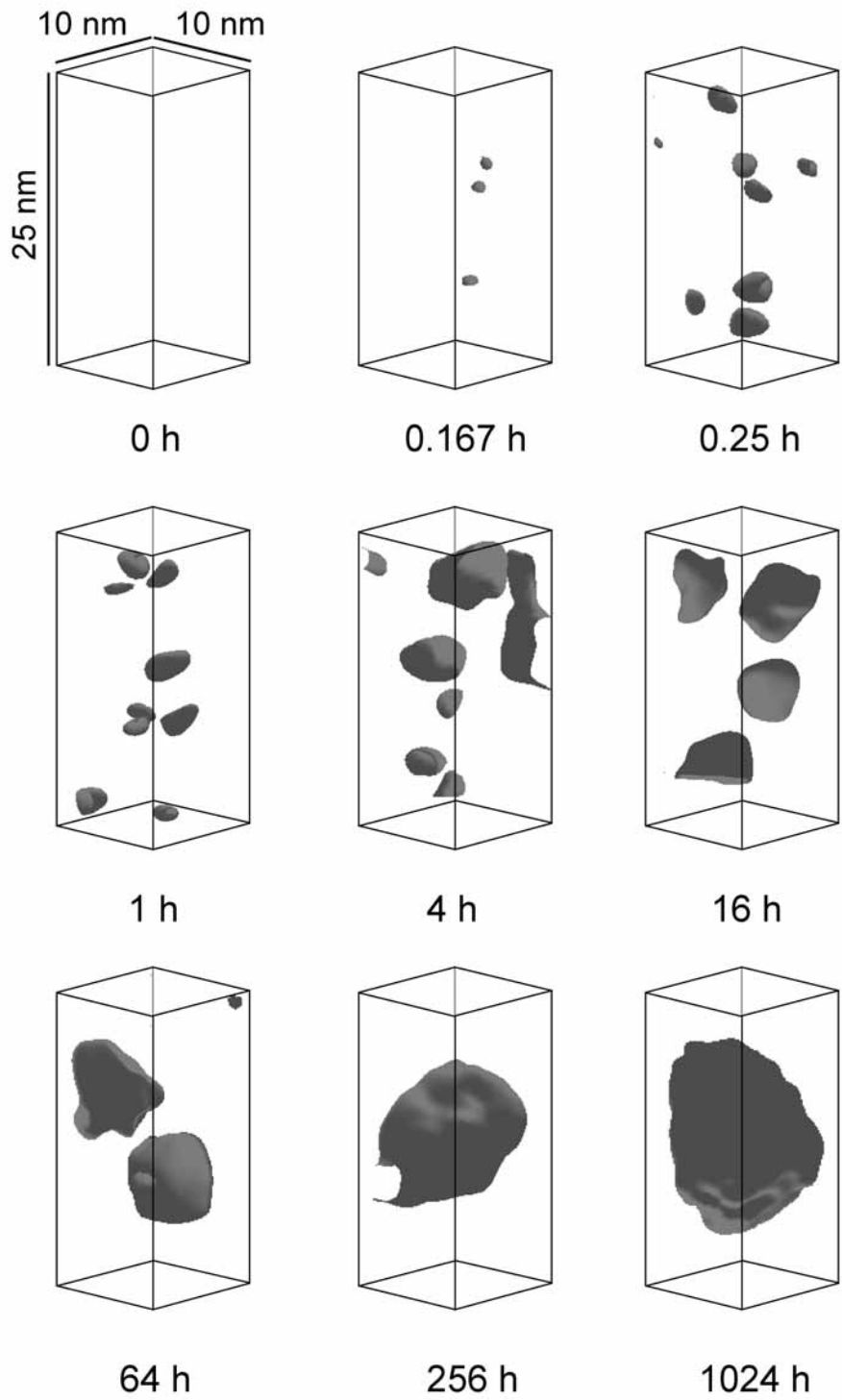

FIGURE 3

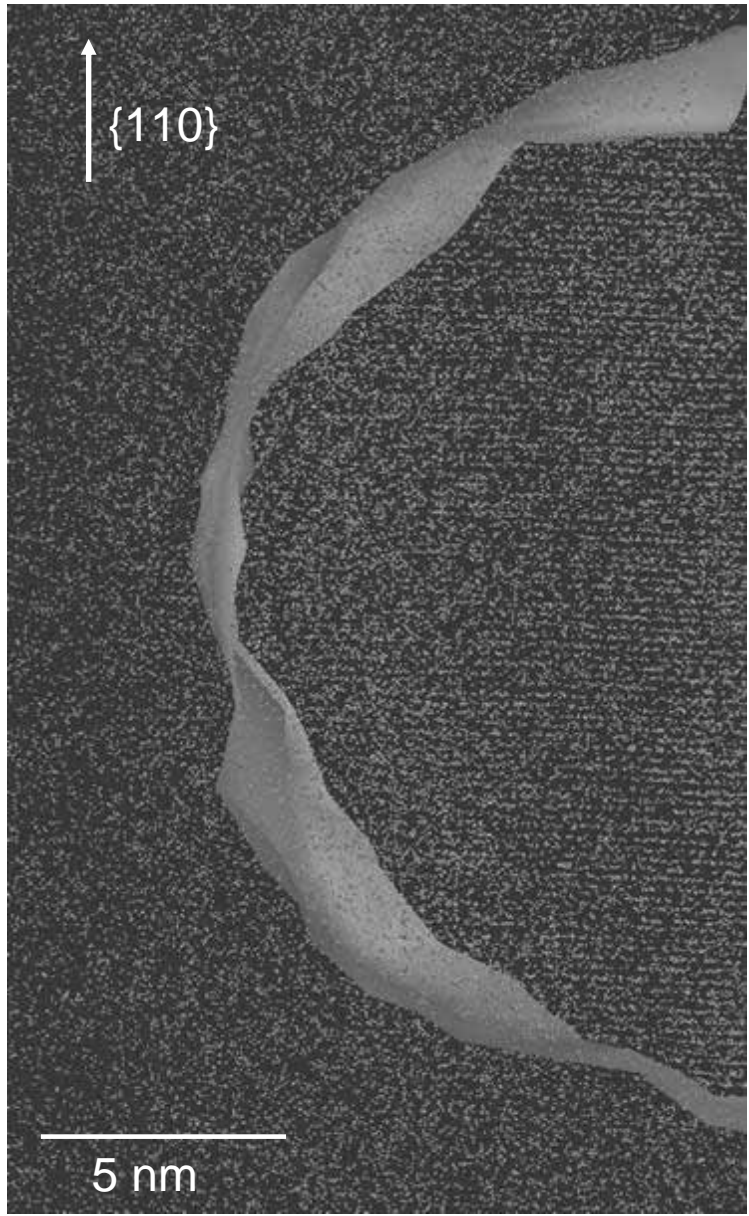

FIGURE 4

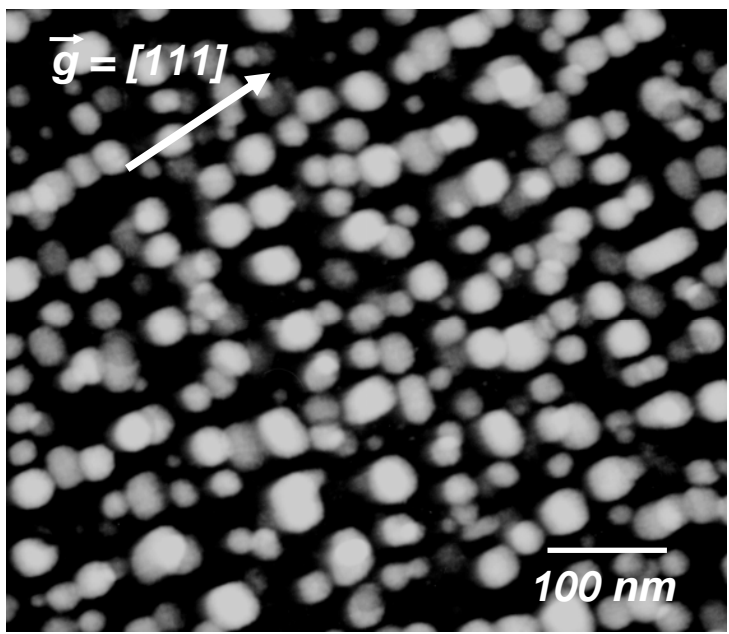

FIGURE 5

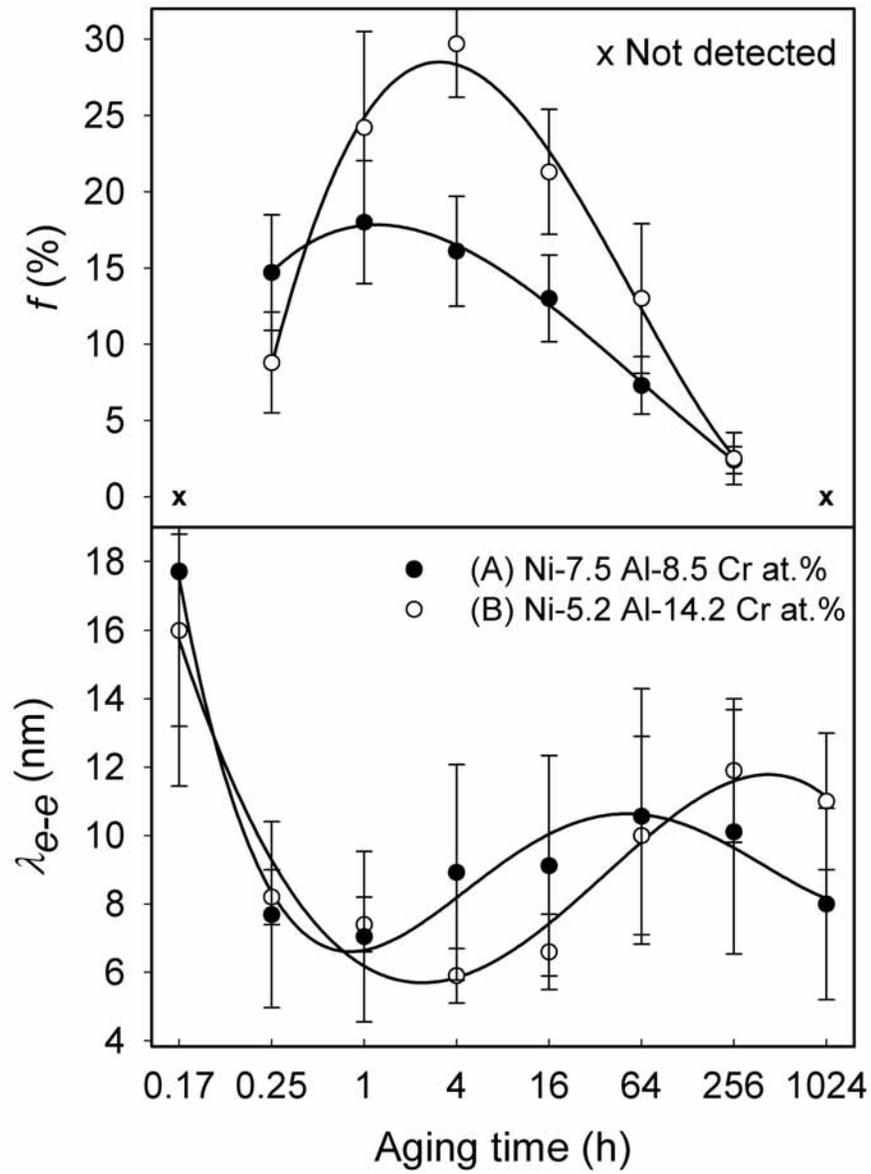

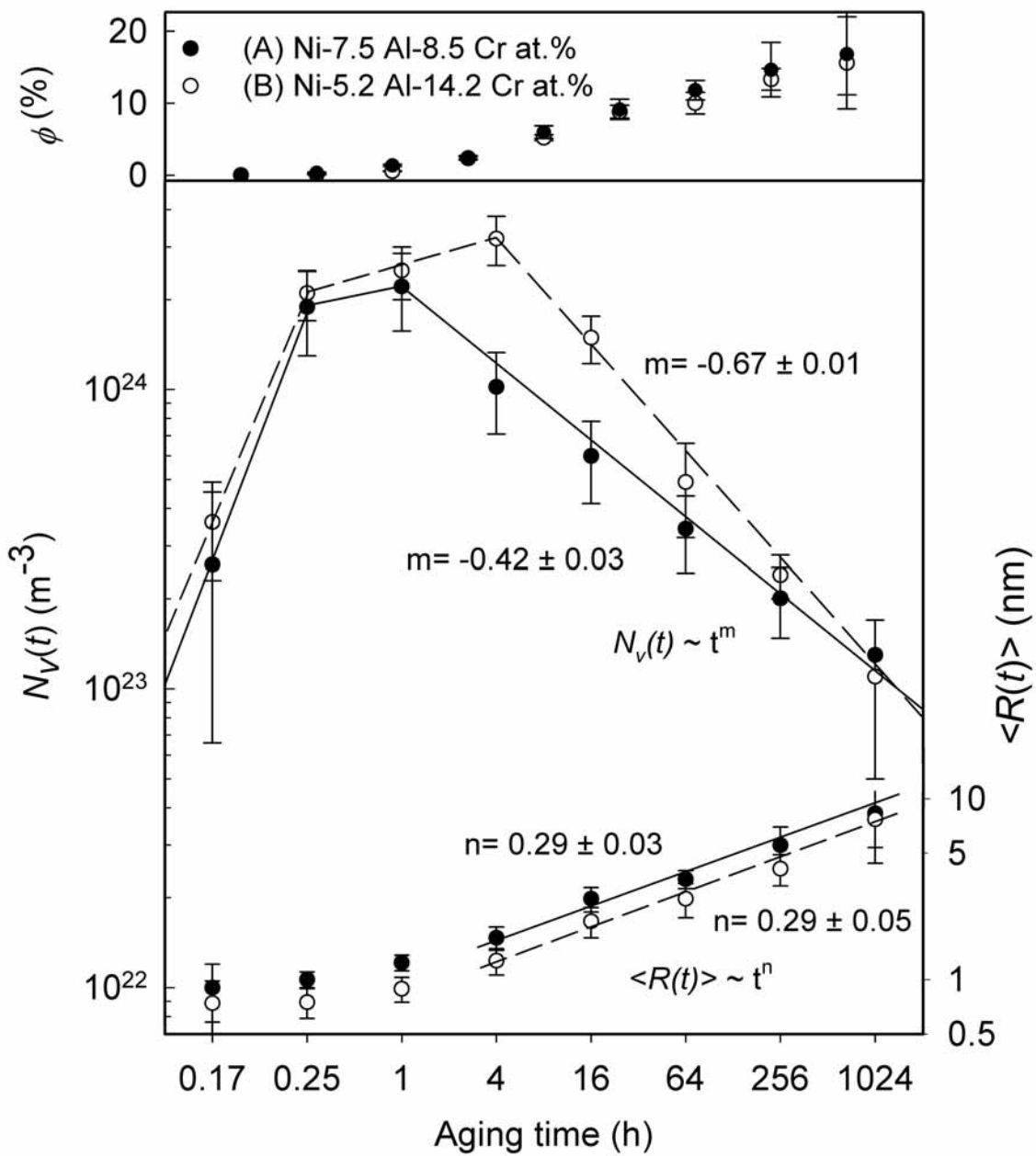

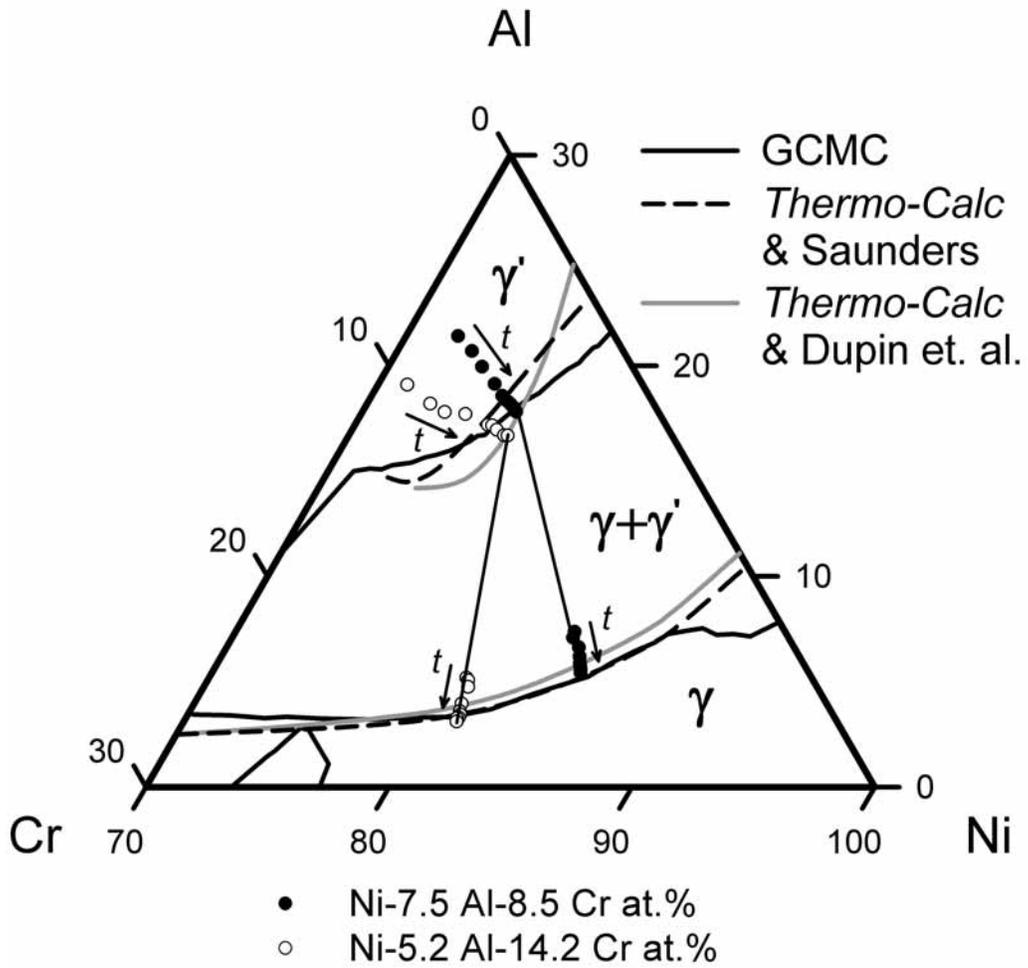

● Ni-7.5 Al-8.5 Cr at.%
○ Ni-5.2 Al-14.2 Cr at.%

FIGURE 8

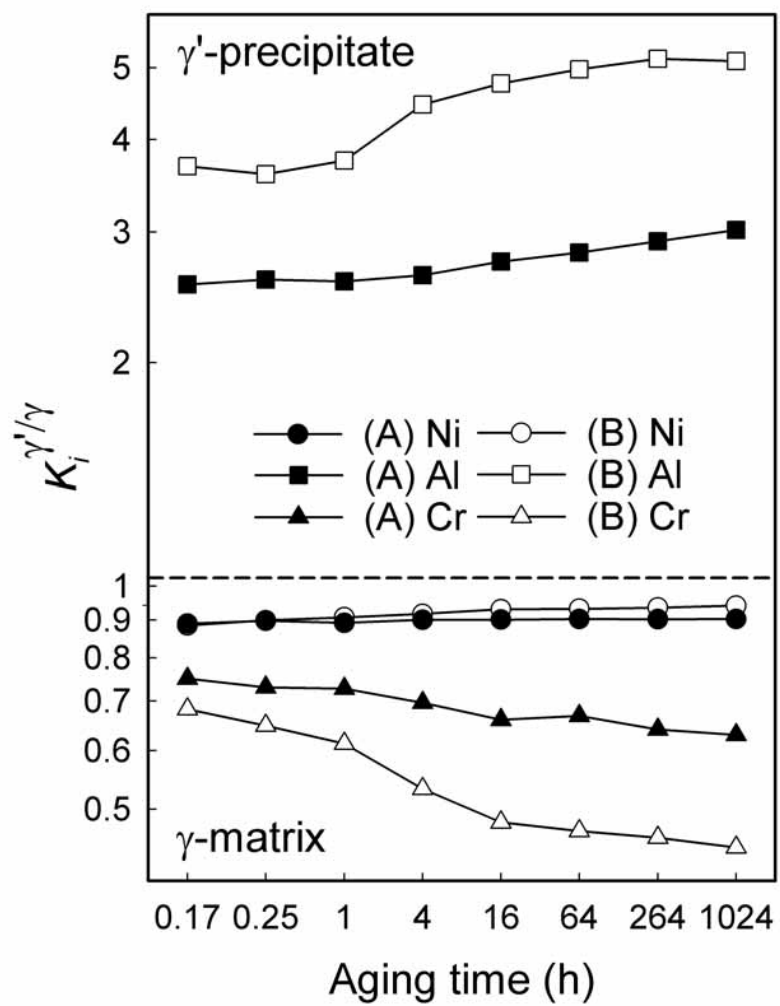

FIGURE 9

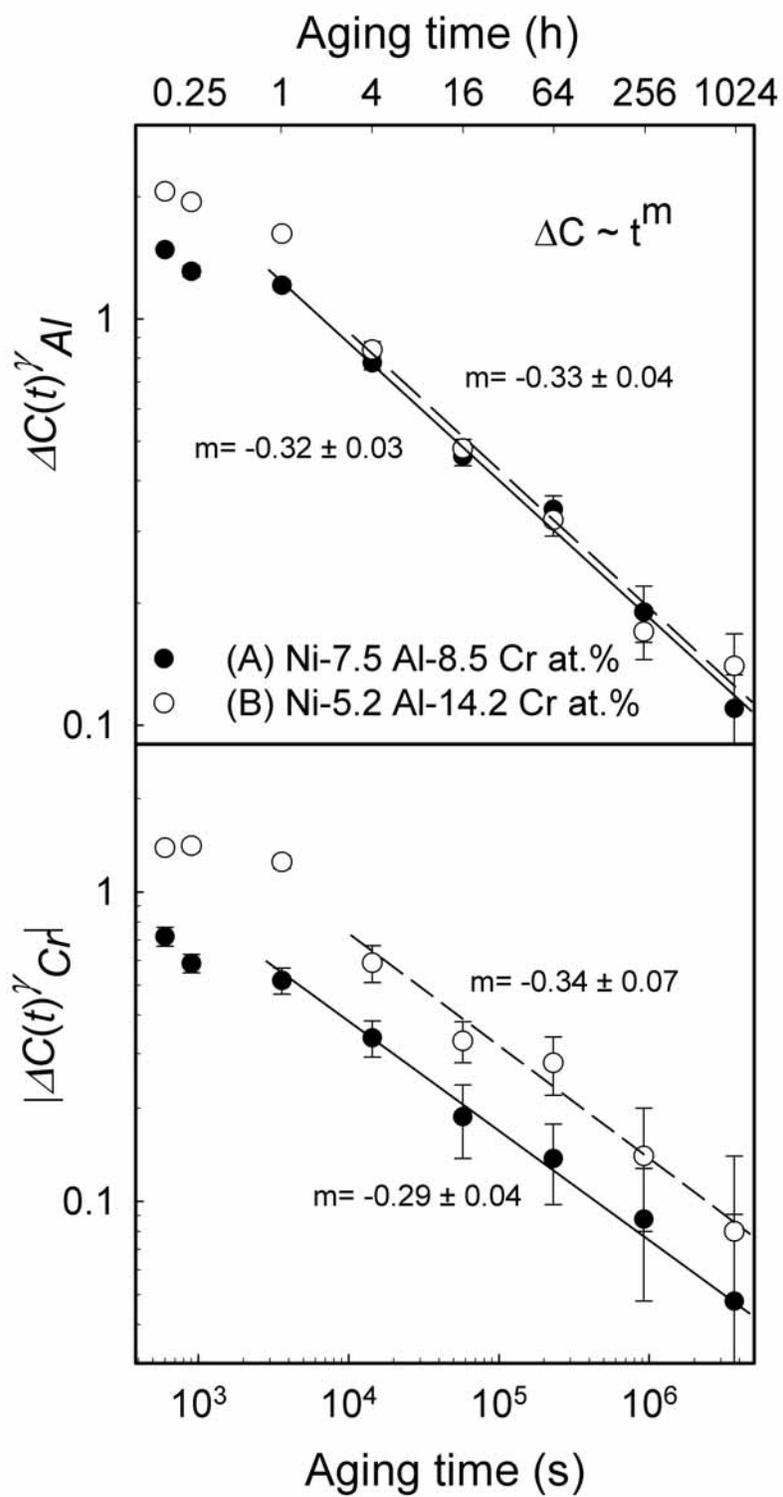

 

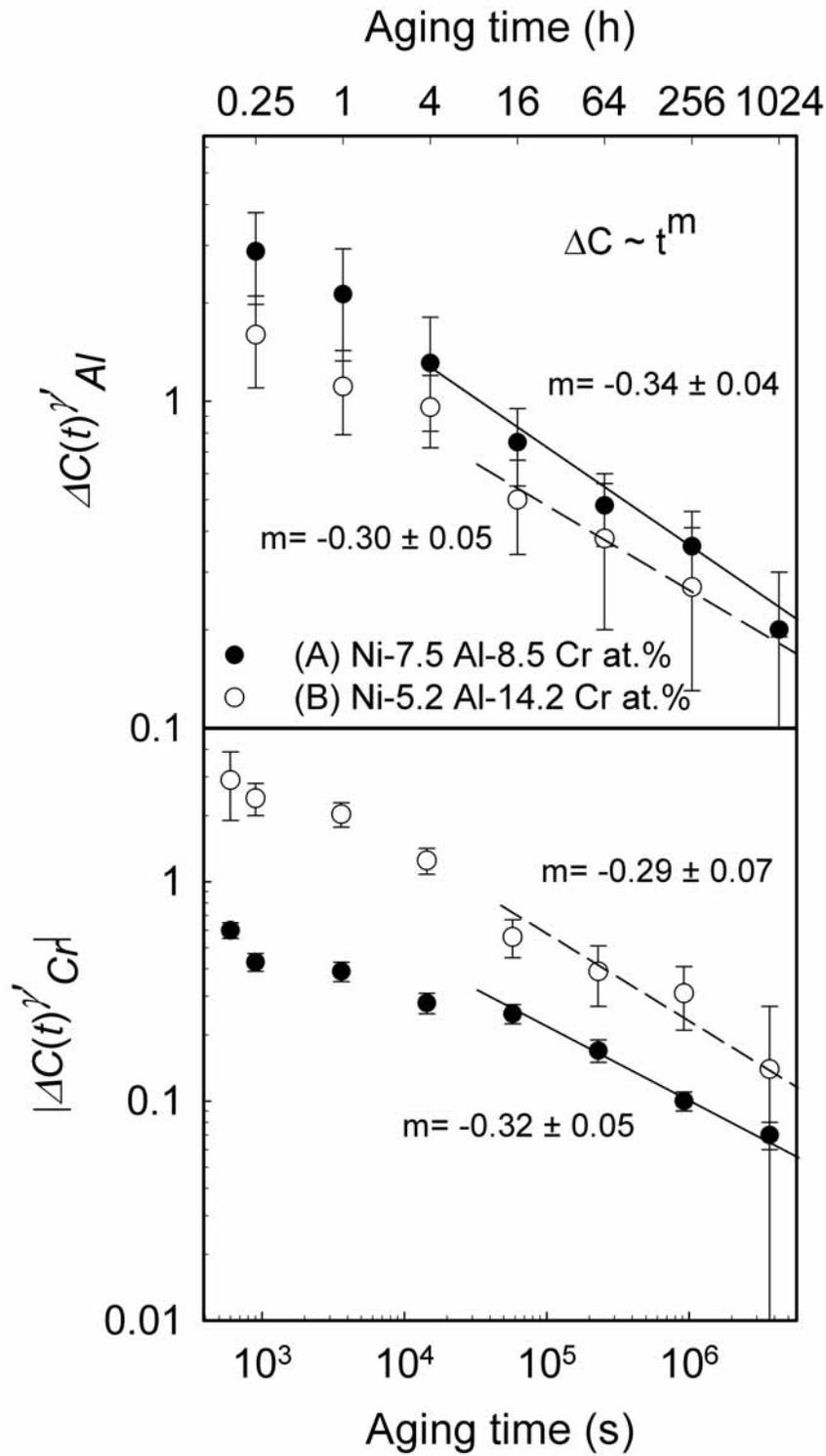